\documentclass[11pt]{amsart}
\allowdisplaybreaks[1]

\usepackage{a4wide}
\usepackage{amssymb}
\usepackage{amsmath}          
\usepackage{amscd} 

\newtheorem{theorem}{Theorem}
\newtheorem{lemma}{Lemma}
\newtheorem{prop}{Proposition}
\newtheorem{coro}{Corollary}

\newtheorem{definition}{Definition}

\newcommand{\ts}{\hspace{0.5pt}}

\newcommand{\CC}{\mathbb{C}\ts}
\newcommand{\RR}{\mathbb{R}\ts}

\newcommand{\ZZ}{\mathbb{Z}}
\newcommand{\NN}{\mathbb{N}}

\newcommand{\TT}{\mathbb{T}}
        
\newcommand{\dd}{\,{\rm d}}

\newcommand{\MM}{\mathcal{M}(G)}
\newcommand{\MCV}{\mathcal{M}_{C,V}(G)}

\newcommand{\Oomega}{(\varOmega,\alpha)}

\newcommand{\LO}{L^2 (\varOmega,m)}

\newcommand{\CalB}{\mathcal{B}}
\newcommand{\Calw}{\mathcal{W}}
\newcommand{\calA}{\mathcal{A}}

\newcommand{\Ghat}{\widehat{G}}
\newcommand{\gammahat}{\widehat{\gamma}}

\newcommand{\axi}{A^{\xi}}
\newcommand{\axib}{A^{\xi}_{B_n}}
\newcommand{\axic}{A^{\xi}_{B_N}}
\newcommand{\axiq}{A^{\xi}_{Q_n}}

\newcommand{\cxi}{c_{\xi}}
\newcommand{\cxib}{c^{\xi}_{B_n}}

\newcommand{\eLL}{\mbox{L}}

\newcommand{\oplam}{\mbox{\Large $\curlywedge$}}

\newcommand{\vL}{\varLambda}

\newcommand{\supp}{\mbox{supp}}

\newcommand{\Sub}{\mbox{Sub}}

\newcommand{\Lperp}{\widetilde{L}^\perp}
\newcommand{\Lnull}{L^\circ}

\newcommand{\cp}{(G,H,\widetilde{L})}
\newcommand{\mcp}{(G,H,\widetilde{L},\rho)}

\newcommand{\symdiff}{\bigtriangleup}

\newcommand{\Hm}[1]{\leavevmode{\marginpar{\tiny%
$\hbox to 0mm{\hspace*{-0.5mm}$\leftarrow$\hss}%
\vcenter{\vrule depth 0.1mm height 0.1mm width \the\marginparwidth}%
\hbox to
0mm{\hss$\rightarrow$\hspace*{-0.5mm}}$\\\relax\raggedright #1}}}

\newcommand{\sigmaprime}{\sigma_\star}

\begin{document}

\title[Continuity of eigenfunctions]
{Continuity of eigenfunctions of uniquely ergodic dynamical systems and
  intensity of Bragg peaks}

\author{Daniel Lenz}
\address{  Current address: Department of Mathematics, Rice University,   P. O. Box 1892, Houston, TX 77251, USA\\ On leave from: Fakult\"at f\"ur Mathematik, D- 09107 Chemnitz, Germany}
\email{dlenz@mathematik.tu-chemnitz.de }
 \urladdr{http://www.tu-chemnitz.de/mathematik/analysis/dlenz}


\begin{abstract}  We study uniquely ergodic dynamical systems over locally
  compact, sigma-compact Abelian groups. We characterize uniform convergence
  in Wiener/Wintner type ergodic theorems in terms of continuity of the limit.
  Our results generalize and unify earlier results of Robinson and Assani
  respectively. 

We then turn to diffraction of quasicrystals and show how the
  Bragg peaks can be calculated via a Wiener/Wintner type result. Combining
  these results we prove a version of what is sometimes known as
  Bombieri/Taylor conjecture. 

Finally, we discuss various examples including
  deformed model sets, percolation models, random displacement models, and
  linearly repetitive systems.
\end{abstract}

\maketitle

\section{Introduction}

This paper is devoted to two related questions. One question concerns
(uniform) convergence in the Wiener/Winter ergodic theorem. The other
question deals with calculating the intensities of Bragg peaks in the
diffraction of quasicrystals and, in particular, with the so called
Bombieri/Taylor conjecture. As shown below the calculation of Bragg
peaks can be reduced to convergence  questions in certain ergodic theorems. The
ergodic theorems of the first part then allow us to prove a version of
the Bombieri/Taylor conjecture.  Let us give an outline of these
topics in this section. More precise statements and definitions will
be given in the subsequent sections.

\smallskip

Consider a topological dynamical system $\Oomega$ over a locally compact,
$\sigma$-compact Abelian group $G$. Let $\xi$ belong to the dual group of $G$
and let $(B_n)$ be a van Hove sequence. We study convergence of averages of
the form
$$ (*)  \:\;\:\;\:\;\:\;\:\;\:\;\:\;\:\;\:\;\:\frac{1}{|B_n|} \int_{B_n} \overline{(\xi,s)}
f(\alpha_{-s}\omega) ds.$$ Due to the von Neumann ergodic theorem and
the Birkhoff ergodic theorem, it is known that these averages converge
in $L^2$ and pointwise to the projection $E_T (\{\xi\})f$ of $f$ on
the eigenspace of $\xi$. In fact (for $G=\RR$ or $G=\ZZ$) the set of
$\omega\in\varOmega$, where pointwise convergence fails, can be chosen
uniformly in $\xi$. This is known as Wiener/Wintner ergodic theorem
after \cite{WW}.

Now, consider a uniquely ergodic $\Oomega$. Unique ergodicity of
$\Oomega$ is equivalent to uniform (in $\omega \in\varOmega$) 
convergence in $(*)$ for $\xi \equiv 1$ and continuous
$f$.  Thus, in this case one might expect uniform convergence for
arbitrary $\xi$ and continuous $f$ on $\varOmega$. The first result of
this paper, Theorem $1$ in Section \ref{Wiener}, characterizes
validity of uniform convergence. It is shown to hold whenever
possible, viz if and only if the limit $E_T (\{\xi\})f$ is continuous.

This generalizes earlier results of Robinson \cite{Rob} for $G=\RR^d$
and $G=\ZZ^d$. More precisely, Robinson's results state uniform
convergence in two situations, viz for continuous eigenvalues $\xi$
and for $\xi$ outside the set of eigenvalues. To us the main
achievement of our result is not so much the generalization of Robinson's
results but rather our new proof. It does not require any
case distinctions but only continuity of the limit. Our line of
argument is related to work of Furman on uniform convergence in
subadditive ergodic theorems \cite{Fur}.

We then study dependence of convergence on $\xi$. Here, again, our result, 
Theorem 2 in Section \ref{Assani}, gives a uniformity statement, provided the limit has strong enough
continuity properties in $\xi$. This result generalizes a result of Assani
\cite{A}, where the limit is identically zero (and thus has the desired
continuity properties). In fact, Theorem 2 unifies the results of Assani and
Robinson. 

\smallskip

While these results are of independent interest, here they serve as tool in the
study of aperiodic order. This is discussed next. 

\smallskip

Aperiodic order is a specific form of (dis)order intermediate between
periodicity and randomness. It has attracted a lot of attention both in
physics and in mathematics in recent years, see e.g. the monographs and
conference proceedings \cite{BMbook,Jan, Mbook, Pat,Sen}. This is due to its
intriguing properties following from its characteristic intermediate form of
(dis)order. In particular, the interest rose substantially after the actual
discovery of physical substances, later called quasicrystals, which exhibit
such a form of (dis)order \cite{dany,INF}.

These solids were discovered in diffraction experiments by their unusual
diffraction patterns. These patterns have, on the one hand, many points,
called Bragg peaks, indicating long range order. On the other hand these
patterns have symmetries incompatible with a lattice structure. Hence, these
systems are not periodic. Put together, these solids exhibit long range
aperiodic order.  Investigation of mathematical diffraction theory is a key
point in the emerging theory of aperiodic order, see the survey articles
\cite{Betal,Hof2, Lag,Len2, Moody06} and references given there.

The main object of diffraction theory is the diffraction measure $\gammahat$
associated to the structure under investigation (see e.g. the book
\cite{Cowley}). This measure describes the outcome of a physical diffraction
experiment. The sharp spots appearing in a diffraction experiment known as
Bragg peaks are then given as the point part of $\gammahat$ and the intensity
of a Bragg peak $\xi$ is given by $\gammahat(\{\xi\})$.  According to these
considerations central problems in mathematical diffraction theory are to
\begin{itemize}
\item prove pure pointedness of $\gammahat$ or at least existence of a
  ``large'' point component of $\gammahat$ for a given structure, 
\item explicitly determine $\xi$ with $\gammahat(\{\xi\})>0$ and calculate
  $\gammahat(\{\xi\})$ for them. 
\end{itemize}
Starting with the work of Hof \cite{Hof}, these two  problems have
been studied intensely over the last two decades for various models (see
references above). The two main classes of models are primitive substitutions
and cut and project models. Prominent examples such as the Fibonacci model or
Penrose tilings belong to both classes \cite{Sen}.

From the very beginning the use of dynamical systems has been a most helpful
tool in these investigations.  The basic idea is to not consider one single
structure but rather to assemble all structures with the ``same'' form of
(dis)order (see e.g. \cite{Rad}). This assembly will be invariant under
translation and thus give rise to a dynamical system.

\smallskip

There is then a result of Dworkin \cite{Dworkin} showing that the
diffraction spectrum is contained in the dynamical spectrum (see as
well the results of van Enter/Mi\c{e}kisz \cite{EM} for closely
related complementary results).  This so called Dworkin argument has
been extended and applied in various contexts.  In particular it has
been the main tool in proving pure point diffraction by deducing it
from pure point dynamical spectrum \cite{Rob2,Hof2,Boris2,Martin}.
Recently it has even been shown that pure point dynamical spectrum is
equivalent to pure point diffraction \cite{LMS,BL,Gouere} and that the
set of eigenvalues is just the group generated by the Bragg peaks
\cite{BL}.

These results can be understood as (at least) partially solving the
first problem mentioned above by relating the set of Bragg peaks to
the eigenvalues of the associated dynamical system.

\smallskip

As for the second problem, the main line of reasoning goes as follows: Let the
structure under investigation be given by a uniformly discrete relatively
dense point set $\varLambda $ in $\RR^d$. For $B\subset \RR^d$ bounded with
non-empty interior and $\xi \in
\RR^d$ define $c_B^\xi (\varLambda):=\frac{1}{|B|} \sum_{x\in \varLambda \cap
  B} \exp (-2\pi i \xi x)$, where $|\cdot|$ denotes Lebesgue measure. Then,
the following should hold
$$ (**)\:\;\:\;\:\;\:\;\:\;\:\;\:\;\:\;\:\;\:
\gammahat(\{\xi\})= \lim_{n\to \infty} | c_{C_n}^\xi (\varLambda)|^2,$$
where $C_n$ denotes the cube around the origin with side length $2n$.
In fact, this is a crucial equality both in numerical simulations and in theoretical considerations. 
It is sometimes discussed under the
heading of ``Bombieri/Taylor conjecture''.  In their work \cite{BT,BT2}
Bombieri/Taylor state (for special one-dimensional systems) that the Bragg
peaks are given by those $\xi$ for which $ \lim_{n\to \infty} c_{C_n}^\xi
(\varLambda) \neq 0.$ They do not give a justification for their statement and
it then became known as their conjecture \cite{Hof,Hof2}.

Since then various works have been devoted to proving existence of
  $\lim_{n\to \infty} c_{C_n}^\xi (\varLambda)$ and rigorously
  justifying the validity of $(* *)$. While the case of general
  uniquely ergodic systems is open, it has been shown by Hof in
  \cite{Hof} that $(* *)$ follows whenever a rather uniform
  convergence of $ c_{B_n}^\xi (\varLambda)$ for van Hove sequences
  $(B_n)$ is known (see the work \cite{AM} for a complementary
  result). This, in turn has been used to obtain validity of $(* *)$
  for model sets \cite{Hof2} and for primitive substitutions \cite{GK}
  (see Theorem 5.1 in \cite{Boris2} for related material as well).

Also, it has been mentioned in various degrees of explicitness
\cite{Hof,Hof2,Boris1,Lag} that this uniform convergence of $c_{B_n}^\xi
(\varLambda)$ follows from or is related to continuity of eigenfunctions due to
Robinson's results \cite{Rob} once one is in a dynamical system setting.  So
far no proofs for these statements seem to have appeared.

These questions are addressed in the second part of the paper. Our results
show that $\gammahat(\{\xi\})$ is related to a specific eigenfunction.  More
precisely, we proceed as follows.

In Section \ref{Diffraction} we first
discuss some  background on diffraction  and then introduce the measure
dynamical setting from Baake/Lenz \cite{BL}. This
setting  has the virtue of embracing the two most common frameworks for the
mathematical modeling of quasicrystals viz the framework of point sets used in
mathematical diffraction theory starting with the work \cite{Hof} and the
framework of bounded functions brought forward in \cite{Bak}.  They are are
both just special cases of the measure approach. Using some material from   \cite{BL} we derive 
Theorem $3$ in Section \ref{Diffraction}. It  shows that
$\gammahat(\{\xi\})$ is the norm square of a certain specific eigenfunction to
$\xi$.  This does not require any ergodicity assumptions and relies solely on
the Stone/von Neumann spectral theorem for unitary representations and
\cite{BL}.  

Depending on whether one has the von Neumann ergodic theorem, or
the Birkhoff ergodic theorem or a uniform Wiener-Wintner type theorem at one's
disposal, one can then calculate $\gammahat(\{\xi\})$ as $L^2$-limit,
almost-sure limit or uniform limit of averages of the form $(*)$. These
averages can then be related to the Fourier type averages in $(**)$ by Lemma
$8$ to give the corresponding $L^2$, almost-sure pointwise and uniform
convergence statement in $(**)$.  This is summarized in Theorem \ref{abstract}
in Section \ref{Conjecture}.  In particular, (b) of Theorem \ref{abstract}
shows that an almost sure justification of Bombieri/Taylor holds in arbitrary
ergodic dynamical systems and does not require any continuity assumptions on
the eigenfunction to $\xi$.  As a consequence we obtain in Corollary $3$ a
variant of the Bombieri/Taylor conjecture valid for arbitrary uniquely ergodic
systems.

\smallskip

These abstract results can be used to reprove validity of $(**)$ for the two
most common models of aperiodic order viz primitive substitution models and
models arising from cut and project schemes (see first remark in Section
\ref{Conjecture}).  More importantly, they can be used to prove validity of
$(**)$ for a variety of new models.  In particular, they allow one to obtain
variants of  the Bombieri/Taylor conjecture for several models arising from
strictly aperiodically ordered ones by some randomization or smearing out
process.

In fact, based on the results of this paper a strong version of the
Bombieri/Taylor conjecture is proven by Lenz/Strungaru \cite{LS} for the class
of deformed model sets earlier studied in \cite{BD,BL2,Gouere} and by
Lenz/Richard \cite{LR} for dense Dirac combs introduced in \cite{R}. These
results are shortly sketched in Section \ref{model}.

Moreover, as shown in Section \ref{random} our abstract results yield almost
sure validity of the Bombieri/Taylor conjecture for both percolation models
and random displacement models based on aperiodic order.  These models are
more realistic in that they take into account defects and thermal motion in
solids respectively.   Percolation models based on aperiodic order were
introduced by Hof \cite{Hof3}. There,  equality of various critical probabilities is shown.  
An extension of Hof's work to graphs together
with an application to random operators is then given in recent work of
M\"uller/Richard \cite{MR}. Random displacement in diffraction for a single
object (rather than a dynamical system) is discussed by Hof in \cite{Hof4}. 
In fact, convergence of diffraction for both
percolation and random displacement models  (and quite some further models) has recently been studied by
K\"ulske \cite{Kuelske, Kuelske2}.  His results give rather 
universal convergence of approximants. However, they deal with a smoothed
version of diffraction.  Thus, they do not seem to give validity of the
Bombieri/Taylor conjecture. In this sense, our results 
complement  the corresponding results of \cite{Kuelske}. We refer
to Section \ref{random} for further details.

In the final section, we study so called linearly repetitive Delone dynamical
systems, introduced by Lagarias/Pleasants in \cite{LP}, and their subshift
counterparts, so called linearly recurrent subshifts, studied e.g.  by Durand
in \cite{Du}.  These examples have attracted particular attention in recent
years and have been brought forward as models for perfectly ordered
quasicrystals in \cite{LP}.  Quite remarkably, continuity of eigenfunctions
fails for these models in general as recently shown in \cite{BDM}.
Nevertheless, we are able to establish validity of $ (**)$ for these models.
More generally, we show that uniform convergence of the modules of the
expressions in $(*)$ holds (while uniform convergence of the expressions
themselves may fail) and this gives validity of $(* *)$ as discussed above.
This is based on the subadditive ergodic theorems from \cite{DL,Len}. Let us
emphasize that convergence in these cases does not hold uniformly in the van
Hove sequences but only for so-called Fisher sequences.

\section{Dynamical systems} \label{Generalities}
Our general framework deals with actions of locally compact Abelian groups on
compact spaces. Thus, we start with some basic notation and facts concerning
these topics. These will be used throughout the paper.

Whenever $X$ is a $\sigma$-compact locally compact space 
(by which we include
the Hausdorff property), the space of continuous functions on
$X$ is denoted by $C(X)$ and the subspace of continuous functions with compact
support by $C_c (X)$. For a bounded function $f$ on $X$, we define the
supremum norm by 
$$\|f\|_\infty :=\sup\{|f(x)| : x\in X\}.$$
Equipped with this norm,  the space $C_K (X)$ of
complex continuous functions  on $X$ with support in the compact set 
$K\subset G$ becomes a complete normed space. 
Then, the space $C_c (G)$ is equipped with the locally convex
limit topology induced by the canonical embedding
$C_K(X)\hookrightarrow C_c (X)$, $K\subset G$, compact.

As $X$ is a topological space, it carries a natural $\sigma$-algebra,
namely the Borel $\sigma$-algebra generated by all closed subsets of
$X$.  The set $\mathcal{M} (X)$ of all complex regular Borel measures
on $G$ can then be identified with the space $C_c (X)^\ast$ of
complex valued, continuous linear functionals on $C_c(G)$. This is justified
by the Riesz-Markov representation theorem, compare \cite[Ch.\ 6.5]{Ped}
for details. The space $\mathcal{M} (X)$
carries the vague topology, i.e., the weakest topology that makes all
functionals $\mu\mapsto   \int_X f \dd\mu  $, $\varphi\in C_c (X)$,
continuous.  The total variation of a measure $\mu \in \mathcal{M} (X)$ 
is denoted by $|\mu|$.

Now fix a  $\sigma$-compact locally compact Abelian (LCA) group $G$. Denote
the  Haar measure  on $G$ by  $\theta_G$. The dual group of $G$ is denoted by
$\Ghat$, and the pairing between a character $\xi \in \Ghat$ and $t
\in G$ is written as $(\xi,t)$.  Whenever $G$ acts on the compact
space $\varOmega$ by a continuous action
\begin{equation*}
   \alpha \! : \; G\times \varOmega \; \longrightarrow \; \varOmega
   \, , \quad (t,\omega) \, \mapsto \, \alpha^{}_{t} (\omega) \, ,
\end{equation*} 
where $G\times \varOmega$ carries the product topology, the pair 
$\Oomega$ is called a {\em topological dynamical system\/} over $G$. We
will often write $\alpha^{}_{t}\ts \omega$ for $\alpha^{}_{t} (\omega)$.
An $\alpha$-invariant probability measure is called {\em ergodic\/} 
if every measurable invariant subset
of $\varOmega$ has measure zero or measure one.  The
dynamical system $\Oomega$ is called {\em uniquely ergodic\/} if there exists
a unique $\alpha$-invariant probability measure. 

We will need two further pieces of notation. A map $\Phi$ between dynamical
system $(\varOmega,\alpha)$ and $(\varOmega',\alpha')$ over $G$ is called a
\textit{$G$ - map} if $\Phi (\alpha_t (\omega)) = \alpha'_t (\Phi (\omega))$
for all $\omega \in \varOmega$ and $t\in G$. A continuous surjective $G$ - map
is called a \textit{factor map}.

\smallskip

Given an $\alpha$-invariant probability measure $m$, we can form the
Hilbert space $\LO$ of square integrable measurable functions on
$\varOmega$. This space is equipped with the inner product
\begin{equation*}
    \langle f, g\rangle \; = \;
    \langle f, g\rangle^{}_\varOmega \; := \; 
    \int_\varOmega \overline{f(\omega)}\, g(\omega) \dd m(\omega).
\end{equation*}
The action $\alpha$ gives rise to a unitary representation $T =
T^{(\varOmega,\alpha,m)}$ of $G$ on $\LO$ by
\begin{equation*}
  T_t \! : \; \LO \; \longrightarrow \; \LO \, , 
  \quad (T_t f) (\omega) \; := \;
  f(\alpha^{}_{-t}\ts \omega) \, ,
\end{equation*}
for every $f\in \LO$ and arbitrary $t\in G$. 
An $f\in \LO$ is called an {\em eigenfunction\/} of $T$ with
{\em eigenvalue\/} $\xi\in \Ghat$ if $T_t f = (\xi, t) f$ for every
$t\in G$.  An eigenfunction (to $\xi$, say) is called {\em continuous\/} 
if it has a continuous representative $f$  with
$$f(\alpha^{}_{-t} \ts \omega) = (\xi,t)\ts f(\omega),\;\,\mbox{for all
$\omega\in\varOmega$  and $t\in G$}.$$
 
By Stone's theorem, compare \cite[Sec.~36D]{Loomis}, there exists a projection
valued measure
\begin{equation*} 
  E_T\! : \; \mbox{Borel sets of $\Ghat$} \; \longrightarrow \;
  \mbox{projections on $\LO$} 
\end{equation*} 
with
\begin{equation} \label{spectralmeasure}
  \langle f, T_t f \rangle \; = \; 
  \int_{\Ghat} (\xi,t) \dd 
  \langle f,E_T(\xi) f\rangle \; := \; 
  \int_{\Ghat} (\xi, t) \dd \rho^{}_f (\xi)\, , 
\end{equation} 
where $\rho^{}_f$ is the measure on $\Ghat$ defined by
$\rho^{}_f (B) := \langle f, E_T (B)f\rangle$.

We will be concerned with averaging procedures along certain sequences. To
do so we define for  $Q,P\subset G$
the $P$ boundary $ \partial^P Q$ of $Q$  by
$$\partial^P Q :=(( P + Q )\setminus Q^\circ ) 
  \cup (( - P +\overline{G\setminus Q})\cap Q)    ),$$
where the bar denotes the closure of a set and the circle denotes the
interior. As  $G$ is $\sigma$-compact, there exists a sequence
$\{ B_n : n\in\NN\}$ of open, relatively compact sets
$B_n \subset G$ with $\overline{B_n}\subset B_{n+1}$, 
$G=\bigcup_{n\ge 1} B_n$,  and 
\[
   \lim_{n\to\infty} \frac{\theta_G (\partial^K B_n) }{\theta(B_n)}=0,
\]
for every compact $K\subset G$ see \cite{Martin} for details. Such a
sequence is called a {\em van Hove sequence\/}.

The relevant averaging  operator is defined next. 

\begin{definition} Let $\Oomega$ be a dynamical system over $G$. For $\xi \in
  \Ghat$,  $B\subset G$ relatively compact  with non-empty interior  and
a  bounded measurable $f$ on $\varOmega$, the bounded measurable function
  $\axi_{B} (f)$ on $\varOmega$  is defined by 
$$ \axi_{B} (f)(\omega) :=\frac{1}{\theta_G (B)} \int_B \overline{(\xi,s)}
f(\alpha_{-s} \omega) ds.$$
In particular,  $ \axi_{B}$  maps $C(\varOmega)$ into itself. 
\end{definition}

In this context the von Neumann ergodic theorem (see e.g. \cite[Thm.\
6.4.1]{Krengel}) gives the following. 

\begin{lemma}\label{vNET}
Let $\Oomega$ be a dynamical system over $G$ with $\alpha$-invariant
probability measure $m$ and associated spectral family $E_T$. Then, 
$$\axib(f) \longrightarrow E_T ( \{\xi\} ) f, \, n\to \infty,$$
in $\LO$ for any $f\in C(\varOmega)$ and every van Hove sequence $(B_n)$. 
\end{lemma}

There is also a corresponding well-known pointwise statement. 

\begin{lemma}\label{pointwiseWWET} 
Let $\Oomega$ be a dynamical system and $m$ an
  ergodic invariant probability measure on $\varOmega$. Let $(B_n)$ be a van
  Hove sequence in $G$ along which the Birkhoff ergodic theorem holds. Then, 
$\axib (f)$  converges almost surely (and in $\LO$) to $E_T (\{\xi\}) f$ for
  every $f\in C(\varOmega)$. 
\end{lemma}   

\textbf{Remark.} As shown by Lindenstrauss in \cite{Lin}, every amenable group
admits a van Hove sequence along which the Birkhoff ergodic theorem holds.

\begin{proof}[Proof of Lemma \ref{pointwiseWWET}] Let $\TT$ be the unit circle and consider the dynamical system
  $\varOmega \times \TT$ with action of $G$ given by $\alpha_s
  (\omega,\theta)) = (\alpha_s \omega, ( \xi, s) \theta)$. Then, the statement
  follows from Birkhoff ergodic theorem applied to $F (\omega, \theta)= \theta
  f(\omega)$.
\end{proof}

In order to prove our abstract results, we need two more  preparatory results.

\begin{lemma}\label{smear} Let $\Oomega$ be a dynamical system over $G$. Let
  $Q,P$ be open, relatively compact non-empty subsets of $G$. 
Then, 
$$ \|\axi_Q (f) - \axi_P ( \axi_Q (f)   )\|_\infty  \leq \frac{\theta_G
  (\partial^{P \cup (-P)} Q)}{\theta_G (Q)} \|f\|_\infty $$
for every $f\in C(\varOmega)$. 
\end{lemma}
\begin{proof} For $t\in G$ a direct calculation shows
\begin{eqnarray*}
|\axi_Q (f)(\omega)  -  \overline{(\xi,t)} \axi_Q (\alpha_{-t}\omega)| 
& = & \frac{1}{\theta_G (Q)} \left| \int_Q  f (\alpha_{-s}\omega) \overline{(\xi,s)}
  ds - \int_{t + Q}  f (\alpha_{-s}\omega) \overline{(\xi,s)}
  ds\right| \\
&\leq &  \frac{ \theta_G (Q \setminus (t +Q) \cup (t+ Q) \setminus Q )
}{\theta_G (Q)} \|f\|_\infty.
\end{eqnarray*}
For $t\in P$, we have  $Q \setminus (t +Q) \cup (t+ Q) \setminus Q
\subset \partial^{P \cup (-P)} Q$ and the lemma follows. 
\end{proof}

The following lemma is certainly well known. We include a proof for
completeness. 

\begin{lemma}\label{halbstetig} Let $\Oomega$ be uniquely ergodic with unique
  $\alpha$-invariant probability measure $m$. Let $K\subset \varOmega$ be
  compact and denote the characteristic function of $K$ by $\chi_K$.  Then,
  for every van Hove sequence $(B_n)$
$$ \limsup_{n\to \infty} \frac{1}{|B_n|} \int_{B_n} \chi_{K}
(\alpha_{-s}\omega) ds \leq m(K)$$
uniformly in $\omega\in \varOmega$. 
\end{lemma}
\begin{proof} As $\varOmega$ is compact, the measure $m$ is regular.  
In particular, for every $\varepsilon>0$, we can find an open set $V$
containing $K$ with $m(V) \leq m(K) +\varepsilon$. By Urysohns lemma, we can
then find   a continuous
  function $h : \varOmega \longrightarrow [0,1]$ with support contained in $V$
  and $h\equiv 1$ on $K$. By construction,   $\int_\varOmega h (\omega) d m
  (\omega) \leq m(K) + \varepsilon$  and 
  $$
  0 \leq \frac{1}{|B_n|} \int_{B_n} \chi_{K} (\alpha_{-s}\omega) ds \leq
  \frac{1}{|B_n|} \int_{B_n} h (\alpha_{-s}\omega) ds$$
  for every $\omega \in
  \varOmega$. By unique ergodicity, $ \frac{1}{|B_n|} \int_{B_n} h
  (\alpha_{-s}\omega) ds $ converges uniformly on $\varOmega$ to
  $\int_\varOmega h(\omega) d m(\omega)$. Putting this together, we obtain
$$  \limsup_{n\to \infty} \frac{1}{|B_n|} \int_{B_n} \chi_{K}
(\alpha_{-s}\omega) ds\leq\int_\varOmega h(\omega) d
m(\omega) \leq m(K) + \varepsilon$$
uniformly on $\varOmega$.  As $\varepsilon>0$ is arbitrary, the statement of
the lemma follows. 
\end{proof}

\section{Uniform Wiener-Wintner type results}\label{Wiener}

In this section we discuss the following theorem.

\begin{theorem}\label{UniformET}
Let $\Oomega$ be a uniquely ergodic dynamical system over $G$ with
$\alpha$-invariant probability measure $m$. Let $\xi\in \Ghat$ and $f\in
C(\varOmega)$ be arbitrary. Then, the following assertions are equivalent:

\begin{itemize}
\item[(i)] The function $E_T (\{\xi\}) f$ has a continuous 
  representative $g$ satisfying $g(\alpha_{-s} \omega) = (\xi,s)\, g(\omega)$
  for every $s\in G$ and $\omega \in \varOmega$. 
\item[(ii)] For some (and then every) van Hove sequence $(B_n)$, the averages
  $\axib (f)$ converge uniformly, that is to say w.r.t. the supremum norm to a
  function $g$.
\end{itemize}

\end{theorem}

\textbf{Remark.} (a) The hard part of the theorem is the implication
$(i)\Longrightarrow (ii)$. Note that $(i)$ comprises three situations:

\begin{itemize}

\item $\xi$ is an eigenvalue of $T$ with a continuous eigenfunction. 

\item $\xi$ is not an eigenvalue at all  (then $g\equiv 0$). 

\item $\xi$ is an eigenvalue and $f$ is perpendicular to the corresponding
  eigenfunctions (then, again, $g\equiv 0$). 
\end{itemize}

(b) Our proof relies on the von Neumann ergodic theorem, Lemma \ref{vNET}, and
unique ergodicity only. Thus, the proof carries immediately over to give a
semigroup version e.g. for actions of $\NN$, as the von Neumann ergodic
theorem is known then.

(c) The statement $(i)\Longrightarrow (ii)$ is given for the first two
situations of (a) separately by Robinson in \cite{Rob} for actions of
$G=\ZZ^d$ and $G= \RR^d$.  Actually, his proof also works for the
third situation.  Our proof is different in this case and works
for all three situations at the same time. Robinson also has a version
for actions of $\NN$. As mentioned in (b), this can be shown by our
method as well.


\bigskip

\textit{Proof of Theorem \ref{UniformET}}:  

\smallskip

(ii)$ \Longrightarrow$ (i): By assumption there exists a van Hove
sequence $(B_n)$ such that the averages $(\axib (f))$ converge
uniformly to a function $g$. As each $\axib(f)$ is continuous, so is
$g$. Moreover, a direct calculation shows
$$ g(\alpha_{-t}\omega) = \lim_{n\to \infty} \axib(f)
(\alpha_{-t}\omega) = \lim_{n\to \infty} (\xi,t) \axi_{ t+ B_n}
(f)(\omega) = (\xi,t) g(\omega)$$ for all $t\in G$ and
$\omega\in\varOmega$. By Lemma \ref{vNET}, $\axib (f)$ converges in
$\LO$ to $E_T (\{\xi\}) f$. Thus, the uniform convergence of the
$\axib (f)$ to $g$ implies $g= E_T (\{\xi\}) f$. This finishes the
proof of this implication.  (The fact that convergence holds for every
van Hove sequence will be proven along the way of (i)
$\Longrightarrow$ (ii).)

\medskip

(i)$ \Longrightarrow$ (ii):  Let $(B_n)$ be an arbitrary van Hove sequence.

Let $\varepsilon>0$ be given. We show that
$$(\sharp) \hspace{4ex} \|\axib (f) - g\|_\infty \leq 3 \ts( 1 + \|f\|_\infty +
\|g\|_\infty) \ts \varepsilon $$
for all sufficiently large $n$. 

\smallskip

By assumption $(i)$ and Lemma \ref{vNET}, $\axib (f)$ converges in $\LO$ to
the continuous function $g$. This implies $ \lim_{n\to \infty} \mu
(\varOmega_n) = 1$ where
$$\varOmega_n :=\{ \omega \in \varOmega : | \axib(f) (\omega) - g(\omega)| <
\varepsilon\}. $$
In particular, for sufficiently large $N$, we have
$$ m(\varOmega_N) \geq 1 - \varepsilon.$$

Fix such an $N$ and set $b:=\axic(f)$.

\smallskip

As both $b$ and $g$ are continuous, the set
$\varOmega_N$ is open. Thus, its complement $K:=\varOmega \setminus
\varOmega_N$  is compact. 
Let $\chi_{\varOmega_N} $ and $\chi_{K}$ be   the characteristic
functions of $\varOmega_N$ and $K$ respectively. Thus, in
particular, 

\begin{itemize}

\item $ \chi_{\varOmega_N} (\omega)  >0$ implies $\omega\in \varOmega_N$,
  i.e. $|b(\omega) -  g(\omega)|< \varepsilon$. 

\item  $ m(K)  \leq  \varepsilon$ (as $m(\varOmega_N) \geq 1 - \varepsilon$). 
\end{itemize}

Note that 
$\axib (b) = \axic (\axib (f))$ by Fubini's theorem and $\axib (g) = g$ by 
the invariance assumption on $g$. Thus, we can estimate

\begin{eqnarray*}
 \|\axib (f) - g\|_\infty  &\leq & \| \axib (f) - \axic (\axib
 (f))\|_\infty + \| \axib (b) - \axib (g)\|_\infty\\
&\leq &  \| \axib (f) - \axic (\axib
 (f))\|_\infty + \| \axib (\chi_{\varOmega_N} (b -g))\|_{\infty} \\
&+& \| \axib ( \chi_K \,  b)  \|_\infty  + \|\axib (\chi_K\,   g) \|_\infty.
\end{eqnarray*}

We estimate the last four terms:

\smallskip

Term 1: By Lemma \ref{smear}, this term can be estimated from above by 
$$  \frac{\theta_G
  (\partial^{B_N \cup (-B_N)} B_n)}{\theta_G (B_n)} \|f\|_\infty.$$
As $(B_n)$ is a van Hove sequence, this term is smaller than $\varepsilon$ for
  sufficiently large $n$.  

\smallskip

Term 2: As $\chi_{\varOmega_N}(\omega) > 0$ implies $|b(\omega) - g
(\omega)| < \varepsilon$, the estimate $\| \chi_{\varOmega_N} (b -
g)\|_\infty< \varepsilon$ holds. This gives $\| \axib
(\chi_{\varOmega_N} (b -g))\|_{\infty} \leq\| \chi_{\varOmega_N} (b -
g)\|_\infty< \varepsilon$ and the second term is smaller than
$\varepsilon$ for every $n\in \NN$.

\smallskip 

Term 3:  A short calculation shows 
$$\| \axib ( \chi_K\, b)\|_\infty \leq \frac{1}{\theta_G (B_n)} \int_{B_n}
\chi_K (\alpha_{-s} \omega) \ts ds\ts \|b\|_\infty .$$
As $K$ is compact with
$m(K) \leq \varepsilon$, we can then infer from Lemma \ref{halbstetig} that
$\frac{1}{\theta_G (B_n)} \int_{B_n} \chi_K (\alpha_{-s} \omega) ds$ is
uniformly bounded by $2 \varepsilon$ for large enough $n$.  As
$\|b\|_\infty\leq \|f\|_\infty$ by definition of $b$, we conclude that the
third term is smaller than $2 \varepsilon \|f\|_\infty$ for sufficiently large
$n$.

\smallskip

Term 4:  Using the same arguments as in the treatment of the third term, we
see that the fourth term can be estimated above by $2 \varepsilon
\|g\|_\infty$ for sufficiently large $n$. 

\medskip

Putting the estimates together we infer $(\sharp)$. \hfill \qedsymbol

\bigskip

We can use the above method of proof to give a proof for the key
technical lemma of \cite{Rob} in our context.

\begin{lemma} 
Let $\Oomega$ be uniquely ergodic. Let $f\in C(\varOmega)$ and $\xi
\in \Ghat$ be arbitrary. Then, $\lim_{n\to \infty} \| \axib\|_\infty =
\sqrt{\langle E_T (\{\xi\}) f, E_T(\{\xi\}) f \rangle}$.
\end{lemma}
\begin{proof} 
For each $\varepsilon >0$ and $N\in \NN$, Lemma \ref{smear} gives 
$$\| \axib (f)\|_\infty \leq \|\axic ( \axib(f))\|_\infty + \varepsilon
\leq \|\axic ( f) \|_\infty + \varepsilon$$ for sufficiently large
$n\in \NN$. This easily shows existence of the limit $\lim_{n\to
\infty} \| \axib (f)\|_\infty$. 

Now, by the von Neumann ergodic theorem, we have $\eLL^2$-convergence
of $\axib (f)$ to $E_T(\{\xi\}) f$. This gives $\eLL^2$-convergence of
$|\axib (f)|$ to $|E_T(\{\xi\}) f|$ and the latter is almost surely
equal to $c := \sqrt{\langle E_T (\{\xi\}) f, E_T(\{\xi\}) f
\rangle}$.  As each $\eLL^2$ converging sequence contains an almost
surely converging subsequence, we infer that $\lim_{n\to \infty} \|
\axib (f)\|_\infty \geq c$.  

It remains to show $\lim_{n\to \infty} \|\axib (f) \|_\infty \leq c$.
Here, we mimic the previous proof: Choose $\varepsilon >0$ arbitrary.
By $\eLL^2$-convergence of $|\axib (f)|$ to the constant function $c$,
we can find an $N\in \NN$ such that $m ( \varOmega_N)\geq 1 -
\varepsilon$, where
$$\varOmega_N :=\{ \omega\in \varOmega : | |\axic (f)(\omega)| - c | <
\varepsilon\}. $$
Note that $\varOmega_N$ is open and hence
$\varOmega\setminus \varOmega_N$ is compact. For $n$ large enough we then have
(see above)
$$\| \axib (f)\|_\infty \leq \|\axic ( f) \|_\infty + \varepsilon \leq \| \axib
(\chi_{\varOmega_N} \axic (f))\|_\infty + \| \axib
(\chi_{\varOmega\setminus\varOmega_N} \axic (f))\|_\infty +
\varepsilon. $$
For $\omega\in \varOmega_N$, we have $| \axic (f)(\omega)| \leq c +
\varepsilon$. Hence, the first term on the right hand side can be
estimated by $c + \varepsilon$. The second term on the right hand side
can be estimated by $\|f\|_\infty \frac{1}{|B_n|} \int_{B_n}
\chi_{\varOmega\setminus\varOmega_N}(\alpha_s \omega) ds$. By Lemma
\ref{halbstetig}, $ \frac{1}{|B_n|} \int_{B_n}
\chi_{\varOmega\setminus\varOmega_N}(\alpha_s \omega) ds$ is smaller
than $ 2\varepsilon = m(\varOmega\setminus \varOmega_N) +  \varepsilon$ for
sufficiently large $n$. This finishes the proof of the lemma.
\end{proof}

\section{Unifying Theorem 1 and a  result of Assani}\label{Assani}

In this section we present the following consequence and in fact
generalization of the hard part of Theorem \ref{UniformET}, which
generalizes a result of Assani as well.
 
\begin{theorem}\label{TheoremAssani}
  Let $\Oomega$ be a uniquely ergodic dynamical system over $G$ with
  $\alpha$-invariant probability measure $m$. Let $f\in C(\varOmega)$
  and $K\subset \widehat{G}$ be given such that, firstly, for every
  $\xi \in K$ the function $E_T (\{\xi\}) f$ is continuous with $ E_T
  (\{\xi\}) f (\alpha_{-s} \omega) = ( \xi,s) E_T (\{\xi\}) f$ for
  all $\omega\in \varOmega$ and $s\in G$ and, secondly, $K
  \longrightarrow C(\varOmega)$, $\xi \mapsto E_T (\{\xi\}) f$, is
  continuous.  Then,
  $$
  \lim_{n\to \infty} \sup_{\xi \in K} \| A_{B_n}^\xi (f) - E_T (\{\xi\})
  f\|_\infty = 0$$
  for every van Hove sequence $(B_n)$.
\end{theorem}

\textbf{Remark.} (a) Certainly the theorem contains the case $K=\{\xi\}$ and
we recover the implication $(i) \Longrightarrow (ii)$ of Theorem
\ref{UniformET}.

(b) The proof of the theorem relies on the previous theorem and
compactness. Thus, again, there is a semigroup version e.g. for actions of
$\NN$.

\bigskip

The theorem has the following immediate  corollary. 

\begin{coro} Let $G$ be discrete and  let $\Oomega$ be a uniquely ergodic dynamical system over $G$ with
  $\alpha$-invariant probability measure $m$. Let $f\in C(\varOmega)$ be given
  such that $E_T (\{\xi\}) f = 0$ for every $\xi \in \widehat{G}$.  Then,
$$
\lim_{n\to \infty} \sup_{\xi \in \widehat{G}} \| A_{B_n}^\xi (f) \|_\infty
= 0$$
for every van Hove sequence $(B_n)$.
\end{coro}

\textbf{Remark.}  For $G=\ZZ$  the corollary was proven by Assani 1993
in an unpublished manuscript. A published proof can be found in his
book \cite{A}.  In fact, the book gives the semigroup version for actions of
$\NN$.


\bigskip

\textit{Proof of Theorem \ref{TheoremAssani}}: Define for each $n\in \NN$ the
function $b_n : K \longrightarrow \RR$, $\xi \mapsto \|A_{B_n}^\xi (f) - E_T
(\{\xi\}) f\|_\infty$. Then, each $b_n$ is continuous by our assumptions and

\begin{equation}\label{pointwise}  b_n(\xi)\longrightarrow 0, \,n\to \infty,\;\: \end{equation}
for each $\xi \in K$ by Theorem \ref{UniformET}.  Moreover, by the invariance
assumption on $E_T (\{\xi\}) f$, we have $A^\xi_{B_n} ( E_T (\{\xi\}) f ) =
E_T (\{\xi\}) f$ for all $n\in \NN$. Thus, Lemma \ref{smear} and direct
arguments give for all $n,N\in \NN$:
\begin{eqnarray*}
b_n (\xi)  &=&  \| A_{B_n}^\xi (f) -  E_T (\{\xi\}) f\|_\infty = \|  A_{B_n}^\xi (f)  -A^\xi_{B_n} (E_T (\{\xi\}) f ) \|_\infty\\
&\leq & \| A_{B_n}^\xi (f) - A^\xi_{B_N} ( A^\xi_{B_n}(f))\|_\infty +\|
A^\xi_{B_n} ( A^\xi_{B_N}(f)) - A^\xi_{B_n} (  E_T (\{\xi\}) f )\|_\infty\\
&\leq & \frac{\theta_G
  (\partial^{B_N \cup (-B_N)} B_n)}{\theta_G (B_n)} \|f\|_\infty + \|  A^\xi_{B_N}(f)
- E_T (\{\xi\}) f\|_\infty\\
& = &  \frac{\theta_G
  (\partial^{B_N \cup (-B_N)} B_n)}{\theta_G (B_n)} \|f\|_\infty  + b_N (\xi).
\end{eqnarray*}
 As $(B_n)$ is a van Hove sequence this easily shows that
the sequence  $(b_n)$ has the following  monotonicity
property: For  each $N\in \NN$ and $\varepsilon >0$, there exists an $n_0 (N,\varepsilon) \in\NN$ with 
\begin{equation}\label{monotone} b_n (\xi) \leq b_N (\xi) + \varepsilon
\end{equation}
for all $n \geq n_0 (N,\varepsilon)$ and all $\xi \in K$. 
Given \eqref{pointwise} and \eqref{monotone},  the theorem follows from
compactness of $K$ and continuity of the $b_n$, $n\in \NN$,  by standard reasoning. \hfill \qedsymbol

\section{Diffraction theory}\label{Diffraction}
In this section we present a basic setup for diffraction (see e.g.
\cite{Cowley}).  For models with aperiodic order this framework has been
advocated by Hof \cite{Hof} and become a standard by now (see introduction for
references). The crucial quantity is a measure, called the diffraction measure
and denoted by $\gammahat$.  It models the outcome of a diffraction experiment
by representing the intensity (per unit volume).  We begin the section with a
short discussion of background material. We then discuss the measure based
approach developed recently \cite{BL,BL2}. We also present a consequence of
\cite{BL} viz Theorem \ref{abstractEigenvalue} (and its corollary), which will
be used in the next section. It may be of independent interest.  We then
finish this section by elaborating on how the usual approach via point sets
fits into the measure approach.

\bigskip

In a diffraction experiment a solid is put into an incoming beam of e.g. $X$
rays. The atoms of the solid then interact with the beam and one obtains an
out coming wave. The intensity of this wave is then measured on a screen.  When
modeling diffraction, the two basic principles are the following: Firstly,
each point $x$ in the solid gives rise to a wave $\xi \mapsto \exp(-i x \xi)$.
The overall wave $w$ is the sum of the single waves. Secondly, the quantity
measured in an experiment is the intensity given as the square of the modulus
of the wave function.

We start with by implementing this for a finite set $F\subset \RR^d$.  Each
$x\in F$ gives rise to a wave $\xi \mapsto \exp(-i x \xi)$ and the  overall
wavefunction $w_F$ induced by $F$ is accordingly
$$ w_F (\xi)  = \sum_{x\in F} \exp(-i x \xi).$$
Thus, the intensity $I_F$ is 
\begin{equation} \label{intensity}
I_F (\xi) = \sum_{x,y\in F} \exp(-i (x-y) \xi) = \widehat{(\sum_{x,y\in F}} \delta_{x-y}).
\end{equation}
Here, $\delta_z$ is the unit point mass at $z$ and ${}\widehat{}$ denotes the
Fourier transform.  When describing diffraction for a solid with many atoms it
is common to model the solid by an infinite set in $\RR^d$.  When trying to
establish a formalism as above for an infinite set $\vL$, one faces the problem that
$$
w_\vL = \sum_{x\in \vL} \exp(-i x\xi)$$
diverges heavily and therefore does
not make sense.  This problem can not be overcome by interpreting the sum as a
tempered distribution.  The reason is that we are actually not interested in
$w_\vL$ but rather in $|w_\vL|^2$. Now, neither modulus nor products are
defined for distributions.  There is a physical reason behind the divergence:
The intensity of the whole set $\vL$ is really infinite. The correct quantity
to consider is not the intensity but a normalized intensity viz  the
intensity per unit volume.  It is given as
$$
I =\lim_{n\to\infty} \frac{1}{|B_n|} I_{\vL\cap B_n}.$$
Of course,  existence of this limit is not clear at all. In fact, we will  even have to
specify in which sense existence of the limit is meant.  It turns out that
existence of the limit in the vague sense is equivalent to existence of the
limit 
$$\gamma= \lim_{n\to \infty} \frac{1}{|B_n|} \sum_{x,y\in \vL\cap B_n} \delta_{x-y}$$
in the vague
sense. In this case, $I$ is the Fourier transform $\gammahat$ of $\gamma$. Then, $\gamma$
is known as autocorrelation function and $I=\gammahat$ is known as diffraction
measure.  We are particularly interested in the point part of $\gammahat$.
The points $\xi\in \RR^d$ with $ \gammahat(\{\xi\})\neq 0$ are called
\textit{Bragg peaks}. The value $\gammahat( \{\xi\} )$ is called the
\textit{intensity of the Bragg peak}.  Particularly relevant questions in this
context are the following:
\begin{itemize}
\item When is $\gammahat$ a pure point measure?
\item Where are the Bragg peaks?
\item What are the intensities of the Bragg peaks?
\end{itemize}
These questions have been discussed in a variety of settings by various people
(see introduction). Here, we  will now present  the framework  and (part of)
the results concerning the first two questions  developed in
\cite{BL,BL2}. The study of the last question is the main content of the
remainder of the paper.

\smallskip

We will be concerned with suitable subsets of the set $\MM$ of measures on
$G$.  There is a canonical map
\begin{equation}\label{deff}
 f : C_c (G)\longrightarrow C(\MM), \;\: f_\varphi (\mu) :=\int_G
\varphi (-s) \,d\mu(s).
\end{equation}
The vague topology on $\MM$ is the smallest
topology which makes all the $f_\varphi$, $\varphi \in C_c (G)$,
continuous.  

Let $C>0$ and a relatively compact open set $V$ in $G$ be given.  A
measure $\mu\in \MM $ is called {\em $(C,V)$-translation bounded\/} if
$|\mu|(t + V) \leq C$ for all $t \in G$.  The set of all
$(C,V)$-translation bounded measures is denoted by $\MCV$. The set
$\MCV$ is a compact Hausdorff space in the vague topology. There is a
canonical map
$$ f : C_c (G)\longrightarrow C(\MCV), \;\: f_\varphi (\mu) :=\int_G \varphi
(-s) \,d\mu(s).$$
Moreover, $G$  acts on $\MCV$ via  a continuous action $\alpha$  given by
\begin{equation*}
   \alpha \! : \; G\times  \MCV \; \longrightarrow \; \MCV 
   \, , \quad (t,\mu) \, \mapsto \, \alpha^{}_t\ts \mu 
   \quad \mbox{with} \quad (\alpha^{}_t \ts \mu) (\varphi)\, := \,
   \int_G \varphi (s + t) d\mu(s).
\end{equation*}

\begin{definition} 
  $\Oomega$ is called a dynamical system on the translation bounded 
  measures on\/ $G$  {\rm (TMDS)} if there exist a constant\/ $C >0$ and 
  a relatively compact open set\/ $V\subset G$ such that\/  
  $\varOmega$ is a closed $\alpha$-invariant subset of\/ $\MCV$.
\end{definition}

Having introduced our models, we can now discuss some key issues
of diffraction theory. Let $\Oomega$ be a TMDS, equipped with an
$\alpha$-invariant probability measure $m$.

Then, there exists a unique measure $\gamma=\gamma_m$ on $G$, called the
\textit{autocorrelation measure\/} (often called Patterson function in
crystallography \cite{Cowley}, though it is a measure in our setting) with
\begin{equation}\label{definitionofgamma}
 \gamma \ast \widetilde{\varphi} \ast {\psi} \,(t) = \;
   \langle f_\varphi, T_t f_\psi \rangle
\end{equation} 
for all $\varphi,\psi \in C_c(G)$ and $t\in G$.  The convolution $\varphi \ast
\psi$ is defined by $(\varphi \ast \psi) (t) = \int \varphi (t- s )
\psi (s) \dd s$. For $\psi \in C_c (G)$ the function $\widetilde{\psi}
\in C_c (G)$ is defined by $\widetilde{\psi} (x) =
\overline{\psi (-x)}$.

By \eqref{definitionofgamma} (applied with $t=0$), the measure $\gamma$ is
positive definite.  Therefore, its Fourier transform exists and is a positive
measure $\gammahat$. It is called the {\em diffraction measure}. As discussed
in the beginning of this section, this measure describes the outcome of a
diffraction experiment.

Taking  Fourier transforms in \eqref{definitionofgamma} and
\eqref{spectralmeasure}  with $f = f_\varphi$, we obtain (see Proposition 7 in
\cite{BL} for details)

\begin{equation}\label{sm}
\rho_{f_\varphi} = |\widehat{\varphi}|^2 \gammahat
\end{equation}
for every $\varphi \in C_c (G)$.  This equation can be used to show that
$\gammahat$ is a pure point measure if and only if $T$ has pure point spectrum
\cite{BL}, see \cite{LMS,Martin} as well.  This equation also lies at the
heart of the following theorem.

\begin{theorem}\label{abstractEigenvalue}
Let $\Oomega$ be a TMDS with an $\alpha$-invariant probability measure $m$ and
associated autocorrelation function $\gamma$. Let $\xi\in\Ghat$ be arbitrary.
Then, there exists a unique $\cxi\in \LO$ with
$$ E_T (\{\xi\}) f_\psi = \widehat{\psi}(\xi) \cxi$$
for every $\psi \in C_c (G)$. The function $\cxi$ satisfies
$\gammahat(\{\xi\}  ) = \langle \cxi, \cxi \rangle$.
\end{theorem}
\begin{proof}
 Uniqueness of such a $\cxi$ is
clear.  Existence  and further properties can be shown as follows:

From \eqref{sm} we obtain by a direct polarization  argument that
$$ (\#) \;\;\; \langle f_\varphi, E_T (B)  f_\psi\rangle = \int \chi_B \,
 \overline{\hat{\varphi}}\,   \widehat{\psi} \, d \gamma$$
for all $\varphi, \psi\in C_c (G)$ and $B\subset \Ghat$ measurable. Here,
 $\chi_B$ denotes the characteristic function of $B$.  Note
 that $E_T (B)$ is a projection and therefore $  \langle f_\varphi, E_T (B)
 f_\psi\rangle =  \langle E_T (B) f_\varphi, E_T (B)  f_\psi\rangle$ for
 arbitrary $\varphi,\psi \in C_c (G)$ and $B\subset \Ghat$ measurable.

\smallskip

Now, choose $\sigma\in C_c (G)$ with $\int_G \sigma(s) ds =1$ and define
$\sigmaprime\in C_c (G)$ by $\sigmaprime :=(\xi,\cdot)
\sigma$. Then, $\widehat{\sigmaprime}(\xi)=1$. Define $c_\xi :=E_T (\{\xi\})
f_{\sigma_\star}$. Then, 
$$ \langle c_\xi, c_\xi\rangle =  \langle E_T (\{\xi\})
f_{\sigma_\star},  E_T (\{\xi\}) f_{\sigma_\star} \rangle = \langle
f_{\sigma_\star},  E_T (\{\xi\}) f_{\sigma_\star}\rangle = \gammahat(\{\xi\})$$
where we used $(\#)$ and $\widehat{\sigmaprime}(\xi)=1$ in the last equality.
Moreover, for arbitrary $\psi \in C_c (G)$ a direct calculation using $(\#)$
and the definition of $c_\xi$
shows
$$ \langle  E_T (\{\xi\}) f_\psi  - \widehat{\psi} c_\xi ,   E_T
(\{\xi\}) f_\psi  - \widehat{\psi} c_\xi\rangle = 0.$$
Thus,  $E_T (\{\xi\}) f_\psi  = \widehat{\psi} c_\xi$ for every $\psi
\in C_c (G)$. 
\end{proof}



\begin{coro} \label{formel} 
  Let $\Oomega$ be a TMDS with an $\alpha$-invariant probability measure $m$
  and associated autocorrelation function $\gamma$.  Let $\mathcal{E}$ be the
  set of eigenvalues of $T$.  Then, the pure point part $\gammahat_{pp}$ of
  $\gammahat$ is given as
  $$
  \gammahat_{pp} =\sum_{\xi\in \mathcal{E}} \langle c_\xi, c_\xi \rangle
  \delta_\xi.$$
  In particular, if $T$ has pure point spectrum then
$\gamma =\sum_{\xi\in \mathcal{E}} \langle c_\xi, c_\xi \rangle \delta_\xi.$

\end{coro}
\begin{proof} 
  The characterizing property of $c_\xi$ given in the previous theorem shows
  that a $\xi$ with $c_\xi\neq 0$ is an eigenvalue with eigenfunction $c_\xi$.
  The formula for the norm of $c_\xi$ in the previous theorem then gives the
  first statement. Now, the second statement follows by noting that \eqref{sm}
  together with pure point spectrum of $T$ implies pure point diffraction (see
  \cite{Martin,LMS,BL} as well)  and, hence, $\gammahat =\gammahat_{pp}$.
\end{proof}

Let us finish this section by discussing how the considerations from the
beginning of this section dealing with point sets and diffraction as a limit
fall into the measure framework.

To do so we first note that it is possible to express $\gamma$ (defined via
the closed formula \eqref{definitionofgamma}) via a limiting
procedure in the ergodic case.  The following holds \cite{BL}.

\begin{theorem}\label{calc} 
  Assume that the locally compact abelian $G$ has a countable base of
  topology. Let $\Oomega$ be a TMDS with ergodic measure $m$ and $(B_n)$ a van
  Hove sequence along which the Birkhoff ergodic theorem holds.  Then,
  $\frac{1}{|B_n|} \omega_{B_n} \ast \widetilde{\omega_{B_n}}$ converges to
  $\gamma_m$ vaguely for $m$-almost every $\omega\in \varOmega$.
\end{theorem}

Next, we show how to consider point sets as measures.   The set of discrete point sets in $G$ will be denoted by
$\mathcal{D}$.  Then,
$$
\delta :\mathcal{D} \longrightarrow \MM, \delta_\varLambda :=\sum_{x\in
  \varLambda} \delta_x,$$
is injective. In this way, $\mathcal{D}$ can and
will be identified with a subset of $\MM$. In particular, it inherits the
vague topology.  A subset $\vL$ of $G$ is called 
\textit{relatively dense} if there exists a compact $C \subset G$ with
$$
G = \bigcup_{x\in \vL} (x + C)$$
and it is called 
\textit{uniformly discrete} if there
exists an open neighbourhood $U\subset G$ of the origin such that
\begin{equation} \label{ud} (x +U) \cap (y+U) =\emptyset\end{equation}
for all $x,y\in \varLambda$ with $x\neq y$.  The set of uniformly discrete
sets satisfying \eqref{ud} is denoted by $\mathcal{D}_U$.  An element $\vL\in
\mathcal{D}_U$ (considered as an element of $\MM$) is in fact translation
bounded.  In particular, we can define the hull of $\varLambda$ as the closure
$$\varOmega(\varLambda):=\overline{ \{ x + \varLambda : x\in G \}}.$$
Then, $\varOmega(\varLambda)$ is a compact  TMDS.

\section{The Bombieri/Taylor conjecture for general systems}\label{Conjecture}

In this section we consider a TMDS $\Oomega$ and ask for existence of certain
Fourier type coefficients. These coefficients will be given as limits of
certain averages. These averages will be defined next. 

\begin{definition}
 Let $\Oomega$ be a TMDS. For $\xi \in
  \Ghat$ and $B\subset G$  relatively  compact with  non-empty interior  the
  function
$ c^{\xi}_B : \varOmega\longrightarrow C(\varOmega)$ is defined by
$$
c^{\xi}_B (\omega) :=\frac{1}{\theta_G (B)} \int_B \overline{(\xi,s)}
d\omega(s).$$
\end{definition}

The conjecture of Bombieri/Taylor was originally phrased in the framework of
point dynamical systems over $\ZZ$, see \cite{BT,BT2,Hof,Hof2}. In our context
a specific version of it  may be reformulated as saying that
$$\gammahat(\{\xi\}) = \lim_{n\to \infty} |\cxib (\omega)|^2, $$
where the
limit has to be taken in a suitable sense (see introduction for further
discussion).

\medskip

Our abstract result reads as follows.

\begin{theorem}\label{abstract}
Let $\Oomega$ be a TMDS with an $\alpha$-invariant probability measure $m$ and
associated autocorrelation function $\gamma$. Let $\xi\in\Ghat$ be arbitrary
and $c_\xi\in \LO$ be given by Theorem \ref{abstractEigenvalue}. 
Then, the following assertions hold:

\smallskip

(a) For every van Hove sequence $(B_n)$ the functions $\cxib$ converge
  in $\LO$ to $\cxi$ and 
  $$
  \gammahat(\{\xi\}) = \langle c_\xi, c_\xi \rangle =\lim_{n\to\infty}
  \langle \cxib, \cxib\rangle.$$

\smallskip

(b) If $m$ is ergodic, the function $|\cxi|^2$
  is almost everywhere equal to $ \gammahat(\{\xi\})$. Then, 
$\cxib$
  converge almost everywhere to $\cxi$, whenever
  $(B_n)$ is a van Hove sequence along which the Birkhoff ergodic theorem
  holds. In  particular,   the functions $|\cxib|^2$
  converge almost everywhere to $ \gammahat(\{\xi\})$.

\smallskip

(c) If $\Oomega$ is uniquely ergodic, then the following assertions are
equivalent: 
\begin{itemize}
\item[(i)] $\cxib$ converges uniformly for one (and then any) van Hove
  sequence.  
\item[(ii)] $\cxi$ is a continuous function
satisfying $\cxi (\alpha_{-s}\omega) = (\xi,s)\cxi (\omega)$ for every $s\in
G$ and $\omega\in \varOmega$. 
\item[(iii)] $\cxi\equiv 0$ or $\xi$ is a continuous eigenvalue. 
\end{itemize}
In these cases $\gammahat(\{\xi\}) = \lim_{n\to \infty} |\cxib|^2 (\omega)$
uniformly on $\varOmega$. 
\end{theorem}

\textbf{Remark.} The parts (a) and (b) of the theorem are new. As discussed in
the introduction, validity of
variants of (c) is hinted at in the literature,  see e.g.
\cite{Hof,Hof2, Boris2,Lag}.  However, so far no proof has been given.  For so
called model sets and primitive substitutions validity of (i) in (c) has been
shown by in \cite{Hof} and \cite{GK} respectively. These proofs do
not use continuity of eigenfunctions. On the other hand, continuity of
eigenfunctions is known for so-called model sets \cite{Martin} and primitive
substitution systems \cite{Boris3,Boris4} (see \cite{Host} for the
one-dimensional situation).  Thus, (c) combined with these results on
continuity of eigenfunctions gives a new proof for the validity of
Bombieri/Taylor conjecture for these systems.

\bigskip

 The  theorem gives immediately  the following corollary, which
 proves one version of the Bombieri/Taylor conjecture.

\begin{coro} \label{BTgeneral} 
  Let $\Oomega$ be a uniquely ergodic TMDS with an $\alpha$-invariant
  probability measure $m$ and associated autocorrelation function $\gamma$.
  Then,
  $$\{\xi \in \widehat{G} : \gammahat(\{\xi\}) >0 \} = \{\xi \in \widehat{G} :
  \cxib\;\mbox{does not converge uniformly to $0$}\}.$$
\end{coro}
\textbf{Remark.} The corollary gives an efficient method to prove
$\gammahat(\{\xi\}) >0$, viz it suffices to show that $\cxib(\omega)$ does not
converge to zero for a single $\omega$. The corollary is less useful in
proving  $\gammahat(\{\xi\}) =0$, as in this case one has to check
uniform convergence to zero. This shortcoming will be addressed for special
systems in the final  section of the paper. 

\begin{coro}  
  Let $\Oomega$ be a uniquely ergodic TMDS with an $\alpha$-invariant
  probability measure $m$ and associated autocorrelation function $\gamma$.
  The following assertions are equivalent:

\begin{itemize}
\item[(i)] $\gammahat$ is a pure point measure and $\cxib$ converges uniformly
  for every $\xi\in \Ghat$. 
\item[(ii)] $\LO$ has an orthonormal basis consisting of continuous
  eigenfunctions. 
\end{itemize}
\end{coro}
\begin{proof} As shown in Theorem 7 of \cite{BL}, $\gammahat$ is a pure point
  measure if and only if $\LO$ has an orthonormal basis consisting of
  eigenfunctions. We are thus left with the statement on continuity of
  eigenfunctions. 

Here, $(ii)\Longrightarrow (i)$ follows from (c) of the
  previous theorem.  The implication $(i)\Longrightarrow (ii)$ follows from
  Theorem $8$ in \cite{BL} applied with $\mathcal{V} :=\overline{Lin\{f_\varphi :
  \varphi \in C_c (G)\}}\subset \LO$ after one notices that assumption (i)
  implies by the previous theorem that all eigenfunctions of the restriction
  of $T$ to $\mathcal{V}$ are continuous. 
\end{proof}

We will give the proof of the theorem  at the end of this section. In order to
do so, we need some preparatory results. 

\begin{lemma}\label{UniformvH} Let $C>0$ and $V\subset G$ be open,
  relatively compact and non-empty. Then, for every compact $K\subset G$ and
  every van Hove sequence $(B_n)$
$$\lim_{n\to \infty} \frac{1}{\theta_G (B_n)} \sup\{|\mu|(\partial^K B_n) :
\mu \in \MCV \} = 0. $$
\end{lemma}
\begin{proof} For a fixed $\mu \in \MCV$, the corresponding statement is shown
  by Schlottmann in Lemma 1.1 of \cite{Martin}. Inspection of the proof shows
  that convergence to zero holds uniformly on $\MCV$ (see \cite{LR} for a
  different proof as well).
\end{proof}

\begin{lemma}\label{Fubini}
Let $\Oomega$ be a TMDS with $\varOmega\subset \MCV$. Then, for every $\varphi
\in C_c (G)$, and $B\subset G$ open relatively compact and non-empty the
estimate
$$
\| \axi_{B} (f_\varphi) - \widehat{\varphi}(\xi) c^{\xi}_B\|_\infty \leq
\frac{ C_\varphi }{\theta_G (B)}\left( \sup\{|\mu|(\partial^{S(\varphi)} B) :
  \mu \in \MCV \} + \theta_G(\partial^{S(\varphi)} B )\right)
$$
holds, where $ S(\varphi) :=\supp (\varphi) \cup (- \supp (\varphi))$ and
$C_\varphi := \int |\varphi| dt + \sup\{\int |\varphi|\,d|\mu| : \mu \in
\MCV\}$.  
\end{lemma}  
\begin{proof} Define $D(\omega):= \theta_G (B)^{-1}  |
 \axi_{B} (f_\varphi)(\omega) - \widehat{\varphi}(\xi)
 c^{\xi}_B(\omega)|$. Then, 
\begin{eqnarray*}
D(\omega) &=&  \frac{1}{\theta_G (B)} \left|  \int_G \int_{B} \overline{(\xi,t)} \varphi (t - r) dt
  d\omega (r) - \int_{B} \int_G \overline{(\xi,t)} \varphi (t -r) dt d
  \omega(r)  \right|\\
&\leq&  \frac{1}{\theta_G (B)}  \left|  \int_{G\setminus B} \int_{B} \overline{(\xi,t)} \varphi (t
  - r) dt d \omega(r)  \right| 
+  \left| \int_{B}  \int_{G\setminus B} \overline{(\xi,t)} \varphi (t -r) dt d
  \omega(r)  \right|.
\end{eqnarray*}
It is then straightforward to estimate the first term by  
$\frac{C_\varphi}{\theta_G (B)} |\omega|(\partial^{S(\varphi)} B)
$ and the second term by $\frac{C\varphi}{\theta_G (B)}  \theta_G(\partial^{S(\varphi)}
  B ) $. 
\end{proof}

The next lemma is the crucial link between Wiener/Wintner type averages and
the Fourier coefficient type averages. It shows that the behavior of $\axib
(f_{\varphi})$ is ``the same'' as the behavior of $\widehat{\varphi}(\xi)
\cxib$ for large $n\in \NN$.

\begin{lemma}\label{asymptotic} 
  Let $\Oomega$ be a TMDS with $\varOmega$ and $(B_n)$ an arbitrary van Hove
  sequence. Let $\xi \in \Ghat$ be arbitrary. Then,
$$ \| \axib (f_{\varphi}) - \widehat{\varphi}(\xi)
\cxib\|_\infty \longrightarrow 0, \, n\longrightarrow \infty$$
for every $\varphi \in C_c (G)$.
\end{lemma}
\begin{proof} This is a direct consequence of  Lemma \ref{Fubini} and Lemma
\ref{UniformvH}. 
\end{proof}

\begin{proof}[Proof of Theorem \ref{abstract}]

Choose $\sigma\in C_c (G)$ with $\int_G \sigma(s) ds =1$ and define
$\sigmaprime\in C_c (G)$ by $\sigmaprime :=(\xi,\cdot)
\sigma$. Then, $\widehat{\sigmaprime}(\xi)=1$ and  according to Theorem
\ref{abstractEigenvalue} we have
$$\gammahat(\{\xi\}) = \langle c_\xi, c_\xi \rangle$$
 with $c_\xi = E_T (\{\xi\}) f_{\sigma_\star}$. This will be used repeatedly
 below.

\smallskip

(a)  From Lemma
\ref{vNET} and definition of $c_\xi$, we infer
$$
\gammahat(\{\xi\}) = \langle c_\xi, c_\xi \rangle = \lim_{n\to \infty}
\langle \axib(f_{\sigma_\star}), \axib(f_{\sigma_\star}) \rangle.$$
Now, the
statement follows from Lemma \ref{asymptotic}.

\smallskip

(b) As $\cxi= E_T (\{\xi\}) f_{\sigmaprime}$, the function $\cxi$ is zero or
an eigenfunction to $\xi$. Thus, for fixed $s\in G$, $\cxi( \alpha_{-s}
\omega) = (\xi,s) \cxi (\omega)$ for almost every $\omega \in \varOmega$. In
particular, the function $|\cxi|^2$ is invariant under $\alpha$ and thus, by
ergodicity, almost surely equal to a constant. As $m$ is a probability measure
this constant is equal to $ \langle \cxi, \cxi\rangle$, which in turn equals
$\gammahat(\{\xi\})$.

The statement on almost sure convergence follows from Lemma \ref{asymptotic}
as $\axib (f_{\sigmaprime})$ almost surely converges according to Lemma
\ref{pointwiseWWET}.

\smallskip

(c) This follows from  Lemma  \ref{asymptotic},  Theorem \ref{UniformET} and
the already shown part. 
\end{proof}

As a by-product of our proof we obtain the following result.

\begin{coro}\label{fouriercoefficient} Let $\gamma$ be the autocorrelation of
  a TMDS $\Oomega$.  Let $(B_n)$ be an arbitrary van Hove sequence.  Let $\xi \in \Ghat$ be given.  Then, 
  $$\widehat{\gamma}(\{\xi\}) = \lim_{n\to \infty} \frac{1}{\theta_G (B_n)}
  \int_{B_n} \overline{(\xi,s)} d\gamma (s)$$
  and, similarly,
  $$\widehat{\gamma}(\{\xi\}) = \lim_{n\to \infty}\gamma \ast \varphi_n
  \ast\widetilde{\varphi_n}(0) $$
  for $ \varphi_n := \frac{1}{\theta_G (B_n)}
  (\xi,\cdot)\, \chi_{B_n}\ast \sigma$, where $\sigma$ is an arbitrary element
  of $ C_c (G)$ satisfying $\int_G \sigma dt =1$.
\end{coro}
\textbf{Remark.}  Results of this type play an important role in the
study of diffraction on $\RR^d$ \cite{Hof, Hof2, Stru}. They do not
seem to be known in the generality of locally compact,
$\sigma$-compact Abelian groups we are dealing with here. They can be
inferred, however, from Theorem 11.4 of \cite{GdeL} whenever
transformability of $\gammahat$ is known. This transformability in
turn seems, however, not to be known in general.

\begin{proof} 
  We only consider the first equation.  The second statement can be shown with
  a similar and in fact simpler proof.
  
  \smallskip

Choose $\sigma\in C_c (G)$ and $\varphi_n\in C_c (G)$ as in the
statement, i.e.  with $\int_G \sigma(s) ds =1$  and $\varphi_n :=
\frac{1}{\theta_G (B_n)} (\xi,\cdot)\, \chi_{B_n}\ast \sigma$. Define
$\sigmaprime\in C_c (G)$ by $\sigmaprime :=(\xi,\cdot)\sigma$. Then a
direct calculation shows
$$\widetilde{\sigmaprime} \ast \varphi_n (t) = (\xi,t) \frac{1}{\theta_G (B_n)}
a_n (t)$$ with
$$
a_n (t) = \int_G \int_G \overline{\sigma (s)} \overline{\sigma (-r)}
\chi_{B_n} (t + s - r) d r d s .$$
Thus, with $K:=\supp(\sigma) -\supp
(\sigma)$, we have $a_n (t) = 0 $ for $t\in G \setminus (B_n \cup \partial^K
B_n)$, $a_n (t) = 1 $ for $t\in B_n\setminus \partial^K B$ and $0\leq |a_n
(t)| \leq 1$ for $t\in \partial^K B$. As $(B_n)$ is a van Hove sequence and
$\gamma$ is translation bounded, we then easily infer from Lemma
\ref{UniformvH} that
$$
\gammahat(\{\xi\}) = \lim_{n\to \infty} \frac{1}{\theta_G (B_n)} \int_{B_n}
\overline{(\xi,s)} d\gamma (s)\;\:\;\mbox{if and only if}\;\:\;
\gammahat(\{\xi\}) = \lim_{n\to \infty} \gamma \ast
\widetilde{\sigmaprime}\ast\varphi_n (0).$$
The latter equality can be shown
as follows: A direct calculation shows $\axib (f_{\sigmaprime}) =
f_{\varphi_n}.$ Thus, \eqref{definitionofgamma}, Lemma \ref{vNET} and Theorem
\ref{abstractEigenvalue} show
$$\gamma \ast \widetilde{\sigmaprime}\ast\varphi_n (0)= \langle f_{\sigmaprime},
f_{\varphi_n}\rangle = \langle f_{\sigmaprime}, \axib
(f_{\sigmaprime})\rangle \longrightarrow \langle f_{\sigmaprime}, E_T
(\{\xi\}) f_{\sigmaprime}\rangle = \gammahat(\{\xi\})$$ 
and the proof
is finished.
\end{proof}

\section{Cut and project models and their relatives}\label{model}
In this section we apply the results of the preceding section to model sets
and some variants thereof.  Model sets were introduced by Meyer in \cite{Mey}
quite before the actual discovery of quasicrystals. A motivation of his work
is the quest for sets with a very lattice-like Fourier expansion theory.  In
fact, model sets can be thought of to provide a very natural generalization of
the concept of a lattice.  Together with primitive substitutions they have
become the most prominent examples of aperiodic order.  Accordingly, they have
received quite some attention. We refer the reader to
\cite{Moody,Moody06,Martin} for background and further references.

\medskip

A \textit{cut and project
scheme} over $G$ consists of a locally compact abelian group $H$,
called the internal space, and a lattice $\widetilde{L}$ in
$G\times H$ such that the canonical projection $\pi : G\times H
\longrightarrow G$ is one-to-one between $\tilde{L}$ and
$L:=\pi(\widetilde{L})$ and the image $\pi_{\rm
int}(\widetilde{L})$ of the canonical projection $\pi_{\rm int} :
G\times H\longrightarrow H$ is dense. Given these properties of
the projections $\pi$ and $\pi^{}_{\rm int}$, one can define the
$\star$-map $(.)^\star\!: L \longrightarrow H$ via $x^\star :=
\big( \pi^{}_{\rm int} \circ (\pi|_L)^{-1}\big) (x)$, where
$(\pi|_L)^{-1} (x) = \pi^{-1}(x)\cap\tilde{L}$, for all $x\in L$.

We summarize the features of a cut- and project scheme in the
following diagram:
\begin{equation*} \label{candp}
\begin{array}{cccccl}
    G & \xleftarrow{\,\;\;\pi\;\;\,} & G\times H &
        \xrightarrow{\;\pi^{}_{\rm int}\;} & H & \\
   \cup & & \cup & & \cup & \hspace*{-2ex} \mbox{\small dense} \\
    L & \xleftarrow{\; 1-1 \;} & \tilde{L} &
        \xrightarrow{\,\;\quad\;\,} & L^\star & \\
   {\scriptstyle \parallel} & & & & {\scriptstyle \parallel} \\
    L & & \hspace*{-38pt}
    \xrightarrow{\hspace*{47pt}\star\hspace*{47pt}}
    \hspace*{-38pt}& & L^\star
\end{array}
\end{equation*}
We will assume that the Haar measures on $G$ and on $H$ are chosen in
such a way that a fundamental domain of $\tilde{L}$ has measure $1$.
Given a cut and project
scheme, we can associate to any $W\subset H$, called the  \textit{window}, the
set
\begin{equation*}
   \oplam(W) \; := \; \{ x\in L : x^\star \in W \}
\end{equation*}
A set of the form $t+ \oplam(W)$ is called \textit{model set} if
the  window $W$ is relatively compact with
nonempty interior. Without loss of generality, we may assume that the
stabilizer of the window,
\begin{equation*} \label{window-shifts}
  H_W \; := \; \{ c\in H : c + W = W \}\ts ,
\end{equation*}
is the trivial subgroup of $H$, i.e., $H_W = \{0\}$. A model set is called
{\em regular\/} if $\partial W$ has Haar measure $0$ in $H$. Any model set
turns out to be uniformly discrete.

A central result on model sets (compare \cite{Moody,Martin} and references
given there) states that regular model sets are pure point diffractive, i.e.
$\gammahat$ is a pure point measure.  In fact, the associated dynamical
system $(\varOmega(\varLambda), \alpha)$ obtained by taking the closure
$\varOmega(\varLambda)$ of $\{t+ \varLambda: t\in G\}$ in $\MM$ is uniquely
ergodic with pure point spectrum with continuous eigenfunctions and the
diffraction measure can be calculated explicitely \cite{Hof,Hof2,Martin}.
Given the material of the previous sections we can easily reproduce the
corresponding results.  This is discussed next.  The underlying idea is that
the dynamical system "almost agrees" (in the sense of being an almost
one-to-one extension) with the so called torus parametrization.

A cut and project scheme gives rise to a dynamical system in the following
way: Define $\TT :=(G \times H) / \widetilde{L}$.  By assumption on $
\widetilde{L}$, $\TT$ is a compact abelian group.  Let
\[ G \times H \longrightarrow \TT, \;\:(t,k)\mapsto [t,k],\]
be the canonical quotient map.   Then, there are canonical  
group homomorphisms 
$$\kappa : H \longrightarrow \TT, \;\:h \mapsto [0,h], \:\;\mbox{and}\:\;
\iota : G\longrightarrow \TT, \;\: t \mapsto [t,0].$$
By the defining properties of a cut and project scheme the homomorphism $\iota$
has dense range as $L^\star$ and the homomorphism $\kappa$ is
injective.  There is an action $\alpha'$ of $G$ on $\TT$ via
\[\alpha' : G\times \TT \longrightarrow\TT,\:\; \alpha'_t([s,k]):= \iota(-t) +  [s,k]= [s - t,k].\]
The dynamical system $(\TT,\alpha')$ is minimal and uniquely ergodic, as
$\iota$ has dense range.  Moreover, it has pure point spectrum. In fact, 
the dual group $\widehat{\TT}$ gives a set of eigenfunctions, which
form a complete orthonormal basis by Peter-Weyl theorem. These eigenfunctions
can be described in terms of characters on $G$ and $H$ via the the dual
lattice $\Lperp$ of $\widetilde{L}$ given by
$$\Lperp :=\{ (k,u) \in\widehat{G}\times \widehat{H} : k(l) u(l^\star)
=1\;\:\;\mbox{for all $(l,l^\star) \in \widetilde{L}$}\}.$$
More precisely,
standard reasoning gives that $\widehat{\TT}$ can naturally be identified with
$\Lperp$. In this identification $(k,u)\in \Lperp$ corresponds to $\xi \in
\widehat{\TT}$ with $\xi([t,h])=k(t) u(h)$. This $\xi$ can then easily be seen
to be an eigenfunction to the eigenvalue $k$. 

It turns out that  $k$ already determines
$\xi$ as will be shown next.  Let $\Lnull$ be the set of all $k\in
\widehat{G}$ for which there exists $u\in \widehat{H}$ with $(k,u)\in \Lperp$.
As $\pi_2 (\widetilde{L})$ is dense in $H$, we  infer that $(k,u),
(k,u')\in \Lperp$ implies $u=u'$. Thus, there exists a unique map $\star :
\Lnull\longrightarrow \widehat{H}$ such that
$$\tau: \Lnull \longrightarrow \Lperp, \,\:\; k\mapsto (k,k^\star),$$
is
bijective.

Having discussed the dynamical behavior of $(\TT,\alpha')$ we now
come to the connection between $\varOmega(\varLambda)$ and $\TT$. This
connection is known under the name of torus parametrization \cite{Martin,
  MS}. In Proposition 7 in \cite{BLM} the following version is given.

\begin{prop}\label{blm-prop}
There exists a continuous $G$-map $\beta : \varOmega(\vL)
\longrightarrow \TT$ such that $\beta(\Gamma) = (t,h) +
\widetilde{L}$ if and only if $t + \oplam (W^\circ - h) \subset
\Gamma \subset t+ \oplam(W - h)$.
\end{prop}

For regular model sets, the Haar measure of the boundary of $W$ is zero. Thus,
the previous proposition shows that the set of points in $\TT$ with more than
one inverse image under $\beta$ has measure zero.  This gives easily (see e.g.
\cite{Martin,BLM}) that $(\varOmega,\alpha)$ inherits unique ergodicity and
pure point spectrum with continuous eigenfunctions and eigenvalues $k\in
\Lnull$ from $\TT$.  By Corollary \ref{formel} the diffraction measure can
then be written as
$$
\gammahat = \sum_{k\in \Lnull} \langle c_k, c_k\rangle \delta_k.$$
It
remains to determine the $c_k$.  By continuity of the eigenfunctions and
Theorem \ref{abstract}, the $c_k$ arise as the uniform limit of the function
$c^k_{B_n}$, where $(B_n)$ is an arbitrary van Hove sequence.  For $k\in
\Lnull$ the calculation of this limit can be performed using a convergence
result for cut and project schemes known as uniform distribution. Using the
uniform distribution result of \cite{Moody2001} one obtains
$$c_k (\varGamma) = \overline{ \tau (k) (\beta(\varGamma))} \int_{W}
(k^\star,y  ) dy$$
for $k\in \Lnull$. 
Putting the previous two equations together we obtain
$$\gammahat = \sum_{k\in \Lnull} A_k \delta_k\:\;\mbox{with}\;\: A_k =\left|\int
_{W} (k^\star,y ) dy\right|^2.$$
We refrain from giving further details here but
refer to the next subsection, where a more general situation is treated.

\subsection{Deformed model sets. }
In this subsection we discuss a special form of perturbation of model sets
leading to deformed model sets. These sets have attracted attention in recent
years \cite{BL2,BD,Gouere}.  Based on the results of the previous sections and
\cite{BL2}, it is possible to calculate diffraction measure and
eigenfunctions. Details are worked out in \cite{LS}. Here, we only sketch the
results.

\medskip

We keep the notation used so far. Let $\varLambda :=\oplam (W)$ be a model set
with a regular window. Let $\varOmega$ be its hull. By the discussion above,
$\varOmega$ is uniquely ergodic with invariant probability measure $m$. Let
now a continuous function $\vartheta : W \longrightarrow G$ be given. This map
gives rise to to the perturbed measure
$$\omega_\vartheta:=\sum_{x\in \varLambda} \delta_{x+\vartheta(x^\ast)}.$$
Denote its hull (in $\MM$) by $\varOmega_\vartheta$. As shown in \cite{BL2}
there exists a unique $G$-invariant continuous map
$$\varPhi_\vartheta :\varOmega\longrightarrow \varOmega_\vartheta$$
with $\varPhi_\vartheta (\varLambda)=\omega_\vartheta$ 
and $\varOmega_\vartheta$ inherits unique ergodicity with pure point spectrum
and continuous eigenfunctions $c_k^\vartheta$ to the eigenvalues $k\in \Lnull$
from $\varOmega$.  Thus, again by Corollary \ref{formel}, $\gammahat^\vartheta$
is a pure point measure which can be written as $ \gammahat^\vartheta =
\sum_{k\in \Lnull} \langle c_k^\vartheta, c_k^\vartheta\rangle \delta_k$.  By
Theorem \ref{abstract}, each $c_k^\vartheta$ a limit of $c^k_{B_n}$.  Using
uniform distribution \cite{Moody2001}, we infer that the limit exists and
equals
$$c^k (\Gamma) = \overline{ \tau(k) (\beta(\Gamma))} \int_{W} (k^\star,y )
\overline{(k,\vartheta(y ))} dy$$
for $k\in \Lnull$. We also obtain that the
limits are identically zero for $k\notin \Lnull$.  Accordingly, we find
$$\gammahat^\vartheta =\sum_{k\in \Lnull} A_k \delta_k,\;\:\mbox{with}\;\: A_k
:= \left|\int_W (k^\star,y) \overline{(k,\vartheta(y))} dy\right|^2.$$
Note
that with $\vartheta\equiv 0$ we regain the case of regular model sets.

\subsection{Cut and project models based on measures}

In this subsection we shortly discuss the measure variant of model sets
studied in \cite{LR} (see \cite{R} as well). Based on the results of
the previous sections, it is possible to calculate diffraction and
eigenfunctions in this case. This is carried out in \cite{LR}. Here, we only
sketch the main ideas.

\begin{definition}\label{def:admiss}
  {\rm (a)} A quadruple $\mcp$ is called a {\em measure cut and project
    scheme} if $\cp$ is a cut and project scheme and $\rho$ is an
  $\widetilde{L}$-invariant
  Borel measure  on $G\times H$. \\
  {\rm (b)} Let $\mcp$ be a measure cut and project scheme. A function $f : H
  \longrightarrow \CC$ is called {\em admissible} if it is measurable, locally
  bounded and for arbitrary $\varepsilon>0$ and $\varphi \in C_c (G)$ there
  exists a compact $Q\subset H$ with
\[ \int_{G\times H} |
\varphi (t+s) f(h + k )| (1- 1_{Q} (h + k)) \dd |\rho|(t,h)\leq \varepsilon \]
for every $(s,k)\in G\times H$, where $1_{Q}$ denotes the characteristic
function of $Q$.
\end{definition}

An example of a measure cut and project scheme is given by a cut and project
scheme  $\cp$ and $\rho:= \delta_{\widetilde L} := \sum_{x\in\widetilde{L}} 
\delta_x$.  It is not 
hard to see that then every Riemann integrable $f : H\longrightarrow \CC$ is 
admissible. In this way regular model sets fall within this framework.

Given a measure cut and project scheme  $\mcp$  with an admissible 
$f$ the map 
\begin{equation*}
\nu_f : C_c (G) \longrightarrow \CC, \:\;\varphi \mapsto \int_{G\times H} 
\varphi (t) f(h)
\dd \rho (t,h), 
\end{equation*}
is a translation bounded measure.  Thus, we can consider its hull
\[\varOmega (\nu_f):=\overline{\{\alpha_t (\nu_f) : t\in G\}}.\]
This hull is a TMDS.  

Assume for the remainder of this section that $f$ is not only admissible but
also continuous. Then, it turns out that $(\varOmega (\nu_f),\alpha)$ is
minimal, uniquely ergodic and has pure point spectrum with continuous
eigenfunctions with set of eigenvalues contained $\Lnull$.  In fact, the map
 $$\mu : \TT \longrightarrow \varOmega(\nu_f), \;\:\mu ([s,k]) (\varphi) =\int
 f(h+k) \varphi (s + t) \dd \rho(t,h)$$
 is a continuous surjective $G$-map and
 $ (\varOmega (\nu_f),\alpha)$ inherits pure point spectrum with continuous
 eigenfunctions from $(\TT,\alpha)$.  Note that in terms of factor maps the
 situation here is somehow opposite to the situation considered in the last
 subsection: The dynamical system in question $(\varOmega (\nu_f),\alpha)$ is
 a factor of the torus and not the other way round!

 By pure point spectrum with eigenvalues contained in $\Lnull$ we can write $
 \gammahat = \sum_{k\in \Lnull} \langle c_k , c_k\rangle \delta_k$ by
 Corollary \ref{formel}. Again, by Theorem \ref{abstract}, the $c_k$ can be
 calculated as a uniform limit. The calculation of the limit requires some
 care. The outcome is
$$ 
c_k (\mu ( [s,h] ) ) = \tau (k) (  [s,h] ) \frac{\rho_\TT (\lambda)}{(m_G\times
  m_H)_\TT (1)}  \int f(u) \, ( k^\star, y) \,\dd y.
$$
Here,  $\rho_\TT$ is the unique measure  on $\TT$
with
\begin{equation*}
\int_{G\times H} g(s,h) \dd\rho(s,h) = \int_\TT \sigma_\xi (g)\dd\rho_\TT
(\xi)
\end{equation*}
for all $g\in C_c(G\times H)$, where 
$\sigma_\xi (g) = 
\sum_{(l,l^\star)\in\widetilde{L}} g(s + l,h + l^\star)$ for $g\in 
C_c(G\times H)$. 
Thus, we end up with 
$$\gammahat= \sum_{k\in \Lnull} A_k \delta_k, \:\; \mbox{with} \:\; A_k
=|\frac{\rho_\TT (\tau (k) ) }{(m_G \times m_H)_\TT (1)} \int_H f(u)
(k^\star ,y) \,\dd y|^2.$$

Note that (at least formally) we regain the formula for regular model sets by
choosing $\rho= \delta_{\widetilde{L}}$ and $f$ to be a characteristic
function.

\smallskip

\textbf{Remark.} As shown in \cite{LR}, the set $\varOmega (\nu_f)$ carries a
natural structure of a compact abelian group and $\mu$ is a continuous group
homomorphism.


\section{Examples with randomness}\label{random}
In this section we study diffraction for randomizations of systems with
aperiodic order. We will be particularly interested in models arising via 
percolation process and models arising via a random displacement (sometimes
also known as Mott type disorder).  Both percolation and random displacement
models can be thought of to give a more realistic description of the solid in
question: Percolation takes into account that defects arise. Random
displacement takes into account the thermal movement of the atoms in the
solid. Our results will show that in these cases validity of a (variant) of
the Bombieri/Taylor conjecture is still true! Note that both models rely on
perturbations via  independent identically distributed random variables. 

As discussed in the introduction, Percolation and Random displacement models
based on aperiodic order have been investigated earlier
\cite{Hof3,MR,Hof4,Kuelske,Kuelske2}.  Here, we would like to emphasize the
work of K\"ulske \cite{Kuelske, Kuelske2}. This work gives strong convergence
statements for approximants of the diffraction measure for rather general
situations containing both percolation and random displacement models. In
fact, \cite{Kuelske2} can even treat situations with non i.i.d. random
variables.  Restricted to our setting this provides convergence for
expressions of the form
$$ \frac{1}{|B_n|} \int I_{\vL\cap B_n} (k) \varphi (k) dk$$
for an arbitrary
but fixed $\varphi$ from the space of Schwartz functions.  As this requires
the smoothing with $\varphi$, it does not seem to give any Bombieri/Taylor
type of convergence statement.  In this sense, our results below provide a
natural complement to his corresponding results of \cite{Kuelske}.

Our construction of the percolation model and the proof of its ergodicity seem
to be new. In fact, we present a unified approach to construction and proof of
ergodicity for percolation models and random displacement models.  This may be
of independent interest.

\medskip

We will be interested in point sets and measures in Euclidean space. Thus, our
group is given as $G=\RR^d$.  The $\sigma$-algebra generated by the vague
topology is called Borel $\sigma$-algebra It can be described as follows. A
cylinder set   is a finite union of sets of the form
$$
\{\mu \in \MM : f_{\varphi_j} (\mu) \in I_j, j=1,\ldots,n\}$$
with $n\in
\NN$, $\varphi_j\in C_c (G)$, and $I_j\subset \CC$ measurable. Here,
$f_\varphi$ is defined in \eqref{deff}.  The support of such a cylinder set
is given as the union of the supports of all functions $\varphi$ involved. In
particular, the support of a cylinder set is always compact. 
\begin{lemma} \label{generate}
  Let $\varOmega\subset \MM$ be compact. Then, the set of cylinder sets in
  $\varOmega$ is an algebra (i.e. closed under taking complements and finite
  intersections) and generates the Borel-$\sigma$-algebra. 
\end{lemma}
\begin{proof} 
The set of cylinder sets is obviously  closed under taking finite
intersections and complements. It generates the Borel-$\sigma$-algebra by its
very definition. 
\end{proof}

\begin{lemma} \label{ergodic}
Let $\varOmega\subset \MM$ be a compact $\alpha$-invariant set and
$m$ an ergodic measure on $(\varOmega,\alpha)$. Let $(\nu^\omega)$
be a family of probability measures on $\MM$ satisfying the following
properties:

\begin{enumerate}
\item $\omega \mapsto \nu^\omega (f)$ is measurable for any nonnegative measurable $f$ on
  $\MM$.
  
\item $\nu^{\alpha_t \omega} (f) = \nu^\omega ( f (\alpha_t \cdot))$ for all
  $t\in G$ and $\omega \in\varOmega$.
  
\item There exists a constant $D>0$ with $\nu^\omega (B \cap C) =\nu^\omega
  (B) \nu^\omega (C)$ whenever $B$ and $C$ are cylinder sets with supports of
  distance bigger than $D$.

\end{enumerate}
Then, the measure $m^{(\nu)}$ on $\MM$ with
$$
m^{(\nu)} (f)= \int_\varOmega \left(\int f (\mu) d \nu^\omega (\mu) \right)
d m (\omega)$$
is ergodic.

\end{lemma}
\begin{proof} 
  The proof is a variant of the well-known argument showing ergodicity (and,
  in fact, strong mixing of the Bernoulli shift).  Let $A$ be a measurable
  $\alpha$ invariant set in $\MM$. Define $ f:\varOmega \longrightarrow
  [0,\infty)$ by $f(\omega):= \nu^\omega (A)$. By the assumptions $(1)$ and $(2)$ on $\nu$ and the invariance of 
  $A$ the function $f$ is invariant and measurable. Hence, by ergodicity of
  $m$, $\nu^\omega (A) = m^{(\nu)} (A)$ for $m$ almost every $\omega\in
  \varOmega$.
  
  By Lemma \ref{generate} the algebra of cylinder sets generates the
  Borel-$\sigma$-algebra. Thus, for any $\varepsilon >0$, there exists a cylinder
  set $B$ with $m^{(\nu)} (A\symdiff B)\leq \varepsilon$.  Here, $\symdiff$
  denotes the symmetric difference.  Let $t\in G$ be arbitrary. Then, triangle
  inequality for symmetric differences and invariance of $A$ give
  $$
  m^{(\nu)} ( B\symdiff \alpha_t B) \leq m^{(\nu)} (B \symdiff A) +
  m^{(\nu)} (A \symdiff \alpha_t B) \leq 2 \varepsilon.$$
  As the cylinder set
  $B$ has compact support we can choose $t\in G$ so that the supports of $B$
  and $\alpha_t B$ have distance at least $D$. Hence, $(3)$ yields
$$\nu^\omega (B\cap \alpha_t B) = \nu^\omega (B) \nu^\omega (\alpha_t B)$$
for all $\omega\in \varOmega$.  Combining these formulas we obtain

$$
2 \varepsilon \geq | m (B) - m (B \cap \alpha_t B)| = \int_\varOmega
\nu^\omega (B) ( 1 - \nu^\omega (\alpha_t B)) d m (\omega).$$
As $m^{(\nu)}
(A\symdiff B)\leq \varepsilon$, we infer
$$
\int_\varOmega \nu^\omega (A) ( 1 - \nu^\omega (\alpha_t B)) d m (\omega)
\leq 3 \varepsilon.$$
As $\nu^\omega (A)=m^{(\nu)} (A)$ almost surely  and $m$ is $\alpha$-invariant this gives
$$ m^{(\nu)} (A) - m^{(\nu)} (A) m^{(\nu)} (B)\leq 3 \varepsilon.$$
As this can be inferred for any $\varepsilon>0$ we obtain
$$ m^{(\nu)} (A) - m^{(\nu)} (A)^2 =0.$$
This shows $m^{(\nu)} (A) =1$ or $m^{(\nu)} (A)=0$. 
\end{proof}

\medskip

\textbf{Remark.} The proof shows that assumption  (3) is stronger than
needed. It suffices, to find to each cylinder set with support  $B$ a $t$ with $\nu^\omega
(B\cap \alpha_t B) = \nu^\omega (B) \nu^\omega (\alpha_t B)$. This type of
condition could be required on arbitrary locally compact abelian groups, which
are not compact.  In fact, even an averaged version of this condition can be
seen to be sufficient.

\medskip

\begin{lemma}\label{lln} 
  Let $(c_n)$ be a bounded sequence of complex numbers and $(X_n)$ a sequence
  of bounded identically distributed independent random variables with
  expectation value $E\in \CC$.  Then,
  $$
  \frac{1}{n} \sum_{j=1}^n (c_j X_j - c_j E) \longrightarrow 0, n\to
  \infty,$$
  almost surely.
\end{lemma}
\begin{proof} 
Without loss of generality we can assume that both $(c_n)$ and $(X_n)$
are real-valued. Now, the statement follows from the boundedness
assumption on $(c_n)$ and $(X_n)$  by Kolmogorov criterion.
\end{proof}

We will now fix an open relatively compact neighborhood of the origin of
$\RR^d$ and consider the set $\mathcal{D}_U$ of all uniformly discrete sets
with "distance" $U$ between different points. We will say that
$(\varOmega,\alpha,m)$ with $\varOmega\subset \mathcal{D}_U$ is a dynamical
system if $\varOmega$ is a compact $\alpha$-invariant subset of
$\mathcal{D}_U$ and $m$ is an $\alpha$-invariant probability measure on
$\varOmega$.

\subsection{Percolation models}
In this section we discuss diffraction for percolation models. 

Fix $p\in (0,1)$ and let $\nu_p$ be the probability measure on $\{0,1\}$ with
$\nu_p (\{1\})=p$.  To $\vL\in \mathcal{D}_U$ we associate the product space
$$ S_\vL^P:=\prod_{x\in \vL} \{0,1\}$$ 
with product measure $\nu_\vL' = \prod_{x\in \vL} \nu_p.$
The map
$$
j_\vL^P : S_\vL^P \longrightarrow \MM, \:j_\vL (s):= \sum_{x\in \vL} s(x)
\delta_x$$
allows one to push $\nu_\vL'$ to a measure $\nu^\vL$ on $\MM$ viz
we define
$$\nu^\vL  (f) := \nu_\vL' (f\circ j_\vL^P).$$

The percolation associated to a dynamical system $(\varOmega, \alpha)$
with  $ \varOmega \subset
\mathcal{D}_U$ is then given by the measure $m^P=m^{(\nu)}$ with
$$
m^P (f) = \int_\varOmega \left(\int f(\mu) d\nu^\vL (\mu)\right) dm
(\vL).$$

The following theorem has been proven in \cite{Hof3} and extended in
\cite{MR}. It also follows  from Lemma \ref{ergodic} above.
\begin{theorem} \label{percet}
The measure  $m^P$ is ergodic with support contained in  $\mathcal{D}_U$. 
\end{theorem}
\begin{proof} 
  It suffices to show that assumptions $(1)$, $(2)$ and $(3)$ of Lemma
  \ref{ergodic} are satisfied.  Validity of $(2)$ is clear. $(3)$ follows as
  $\nu_\vL'$ is a product measure.  To show $(1)$ it suffices to show that
  $$
  \vL \mapsto \nu^\vL (f_{\varphi_1}\ldots f_{\varphi_n})$$
  is continuous
  (and hence measurable) for any $n\in \NN$ and $\varphi_1,\ldots,\varphi_n\in
  C_c (G)$. Choose an open relatively compact set $U$ with $U= -U$ containing
  the supports of all $\varphi_j$, $j=1,\ldots,n$. For $s\in \{0,1\}^{\vL \cap
    U}$ let $\sharp_1 s$ and $\sharp_0 s$ denote the number of $1$'s and $0$'s
  in $s$ respectively. Then, a short calculation gives
$$
\nu^\vL ( f_{\varphi_1}\ldots f_{\varphi_n}) = \sum_{ s\in \{0,1\}^{\vL
    \cap U} } \left( \sum_{x\in \vL \cap U} s(x) \varphi_1 (-x)\right) \cdots
\left( \sum_{x\in \vL \cap U} s(x) \varphi_n (-x)\right) p^{\sharp_1 s}
(1-p)^{\sharp_0 (s)}.$$
This easily shows the desired continuity.
\end{proof}

We now turn to diffraction.  The autocorrelation of $\gamma^P=\gamma_{m^P}$
can easily be calculated and seen to be $ \gamma^P = p^2 \gamma_m + p (1 -p)
\delta_0.$ In particular,
\begin{equation} \label{P}
\gammahat^P= p^2 \gammahat_m + p (1 -p)\, 1
\end{equation} 
contains an absolutely continuous component.

\begin{lemma}
  Let $\vL\in \mathcal{D}_U$ and $\xi\in \Ghat$ be given. Let $(B_n)$ be a van
  Hove sequence in $G$. If $c^\xi_{B_n} (\vL)$ converge to a complex number
  $A$, then $c^\xi_{B_n} (\omega)$ converges to $p A$ for $\nu^\vL$ almost
  every $\omega\in \mathcal{D}_U$.
\end{lemma}
\begin{proof}
  It suffices to consider $\omega$ of the form $\omega=j_\vL (s)$ with $s\in
  S_\vL^P$. The lemma then claims that
  $$
  c^\xi_{B_n} (\omega)=\frac{1}{|B_n|} \sum_{x\in \vL \cap B_n} s(x)
  (\xi,x)\longrightarrow p \lim_{n\to \infty} \frac{1}{|B_n|} \sum_{x\in \vL
    \cap B_n}(\xi,x), n\to \infty.$$
  By uniform discreteness of $\vL$, the
  sequence $\frac{\sharp B_n \cap \vL}{|B_n|}$ is bounded. Thus, it suffices
  to show that
  $$
  \frac{1}{\sharp B_n \cap \vL} \sum_{x\in \vL \cap B_n} ( s(x) -
  p)(\xi,x)) \longrightarrow 0, n\to \infty.$$
  This in turn follows easily
  from Lemma \ref{lln}.
\end{proof}

Putting these results together we obtain the following variant of
Bombieri/Taylor conjecture.

\begin{theorem} 
  Let $(\varOmega,\alpha,m)$ with $\varOmega \subset \mathcal{D}_U$ be a
  uniquely ergodic dynamical system with continuous eigenfunctions.  Then, for
  any $\xi\in \Ghat$, and $\vL\in \varOmega$ the averages $c^\xi_{B_n}
  (\omega)$ converge for $\nu^\vL$ almost every $\omega\in \mathcal{D}_U$ to a
  limit $c^\xi (\vL)$. This limit depends only on $\vL$ (and not on $\omega$)
  and satisfies $|c^\xi (\vL)|^2 = p^2 \gammahat_m (\{\xi\})= \gammahat^P (\{
  \xi \} )$. In particular, $|c^\xi_{B_n} (\omega)|^2$ converge for $m^P$
  almost every $\omega\in \varOmega$ to $ \gammahat^P (\{ \xi \} )$.
\end{theorem}
\begin{proof} 
  As $(\varOmega,\alpha)$ is uniquely ergodic with continuous eigenfunctions,
  Theorem \ref{abstract} gives convergence of $c^\xi_{B_n} (\vL)$ to
  continuous functions $c_\xi (\vL) $ with $\gammahat_m (\{\xi\})\equiv
  |c_\xi (\vL) |^2$ for all $\vL\in \varOmega$ and $\xi\in \Ghat$.  The previous
  lemma then proves the $\nu^\vL$ almost sure convergence of $c^\xi_{B_n}
  (\omega)$  for each $\vL$. This lemma and the explicit formula \eqref{P}
  then show the last statement.
\end{proof}

\subsection{Random displacement models}
In this subsection we consider a random displacement in $G=\RR^d$. 
Fix a probability measure $\sigma$ on $\RR^d$ with bounded range. To $\vL\in
\mathcal{D}_U$ we associate the space
$$ S_\vL^{RD}=\prod_{x\in \vL} \RR^d$$
with product measure $\sigma_\vL '= \prod_{x\in \vL} \sigma$. 
The map
$$
j_\vL^{RD} : S_\vL^{RD}\longrightarrow \MM, \:j_\vL^{RD} (s):= \sum_{x\in
  \vL} \delta_{x+ s(x)}$$
allows one to push $\sigma_\vL '$ to a measure
$\sigma^\vL$ on $\MM$ viz we define
$$\sigma^\vL (f) := \sigma_\vL ' (f\circ j_\vL^{RD} ).$$
The random
displacement model associated to a dynamical system $(\varOmega, \alpha,m)$
with $ \varOmega \subset \mathcal{D}_U$, is then given by the measure $m^{RD}
=m^{(\sigma)}$ with
$$
m^{RD} (f) = \int_\varOmega \left(\int f(\mu) d\sigma^\vL (\mu)\right) dm
(\vL).
$$
As $\varOmega$ is compact and $\sigma$ is bounded, the support of $m^{RD}$
is compact.  The following is a consequence of Lemma \ref{ergodic} above. It
seems to be new. The proof is very similar to the proof of Theorem
\ref{percet}.  We omit the details.

\begin{theorem} 
The  measure $m^{RD}$  is  ergodic.  
\end{theorem}

We now turn to diffraction.  By ergodicity and Theorem \ref{calc}, the
autocorrelation $\gamma^{RD}= \gamma_{m^{RD}}$ can be calculated as a limit
almost surely. This limit  has been calculated  in \cite{Hof4}
for a fixed $\vL$ 
and shown to be
$$
\gamma^{RD} = \gamma_m\ast \sigma \ast \widetilde{\sigma} + n_0(\delta_0 -
\sigma\ast \widetilde{\sigma}), $$
where $n_0$ is the density of points. In
particular,
\begin{equation}\label{RD} \gammahat^{RD}= |\widehat{\sigma}|^2 \gammahat_m + n_0 (1 - |\widehat{\sigma}|^2)
\end{equation} 
contains an absolutely continuous component.

The next lemma and the following theorem can now be proven along very similar
lines as the corresponding results in the previous subsection. We omit the
details. 
\begin{lemma}
  Let $\vL\in \mathcal{D}_U$ and $\xi\in \Ghat$ be given. Let $(B_n)$ be a van
  Hove sequence in $G$. If $c^\xi_{B_n} (\vL)$ converge to a complex number
  $A$, then $c^\xi_{B_n} (\omega)$ converges to $\widehat{\sigma} (\xi) A $
  for $\sigma^\vL$ almost every $\omega\in \MM$.
\end{lemma}
\begin{proof}
\begin{theorem} 
  Let $(\varOmega,\alpha,m)$ with $\varOmega\subset \mathcal{D}_U$ be a
  uniquely ergodic dynamical system   with
  continuous eigenfunctions. Then, for any $\xi\in \Ghat$, and $\vL\in
  \varOmega$ the averages $c^\xi_{B_n} (\omega)$ converge for $\sigma^\vL$
  almost every $\omega\in \MM$ to a limit $c^\xi (\vL)$. This limit depends
  only on $\vL$ (and not on $\omega$) and satisfies $|c^\xi (\vL)|^2 =
  \gammahat_m (\{\xi\}) |\widehat{\sigma} (\xi)|^2 = \gammahat^{RD} (\{ \xi \}
  )$. In particular, $|c^\xi_{B_n} (\omega)|^2$ converge for $m^{RD}$ almost
  every $\omega\in \varOmega$ to $ \gammahat^{RD} (\{ \xi \} )$.
\end{theorem}
\end{proof}

\textbf{Remark.}  The above considerations rely essentially on the independent
identical distribution of the randomness and the locallity of the
randomness. Therefore, various  further models can be treated by the same line
of reasoning. In particular, we could   treat models combining  random
displacement with percolation.


\section{Linearly repetitive systems}
In this section we discuss linearly repetitive Delone dynamical systems and
their subshift counterparts known as linearly recurrent subshifts. We will
refer to both classes as LR-systems.
Such systems were introduced recently in \cite{Du,LP} and have further been
studied e.g. in  \cite{CDHM,BDM,DL}. In fact, LR-systems
are brought  as models for
perfectly ordered quasicrystals \cite{LP}. Thus, validity of the Bombieri/Taylor
conjecture for these systems is a rather relevant issue.

LR-systems can be thought of as generalized primitive substitution
systems \cite{Du}. As continuity of eigenfunctions is known for
primitive substitutions \cite{Boris3,Boris4}, it is natural to assume
that continuity holds for LR-systems as well. Somewhat surprisingly,
this turns out to be wrong as discussed in \cite{BDM}. Thus, validity
of the Bombieri/Taylor conjecture can not be derived from the material
presented so far for these models. 

It is nevertheless true as shown below. More generally, we show that
for these systems the modules $|\axib(f)|$ converge uniformly (while
the averages themselves may not converge).  The key to these results
are the uniform subadditive ergodic theorems from \cite{DL,Len}. Let
us caution the reader that these results do not hold for arbitrary van
Hove sequences $(B_n)$ but rather only for Fisher sequences.

We focus on linearly repetitive Delone dynamical systems in this
section and only shortly sketch the subshift case.

\medskip

A subset of $\RR^d$ is called \textit{Delone set} if it is
uniformly discrete and relatively dense (see end of Section \ref{Diffraction} for definition of these notions). 
As usual we will identify a uniformly
discrete subset of $\RR^d$ with the associated translation bounded measure and
this will allow us to speak about e.g. the hull $\varOmega (\vL)$ of a Delone
set.

\begin{definition} The open ball with radius $R$ around the origin is denoted by $U_R (0)$. 
A Delone set $\varLambda$ is called linearly repetitive if
  there exists a $C>0$ such that for all $R\geq 1$, $x\in \RR^d$ and $y\in
  \varLambda$, there exists a $z \in U_{R C}(x) \cap \varLambda$ with
$$ (- z + \varLambda)\cap U_R (0) = (-y + \varLambda) \cap U_R (0).$$
\end{definition}

Roughly speaking, linear repetitivity means that a local configuration
of of size $R$ can be found in any ball of size $C R$.  If
$\varLambda$ is linearly repetitive, then $\varOmega(\varLambda)$ is
minimal and uniquely ergodic.

Linearly repetitive systems allow for a uniform subadditive ergodic
theorem and this will be crucial to our considerations.  The necessary
details are given next. A subset of $\RR^d$ of the form
$$ I_1\times \ldots \times I_d,$$ with nonempty bounded intervals $I_j$,
$j=1,\ldots, d$, of $\RR$ is called a box.  The lengths of the
intervals $I_j$, $j=1, \ldots, d$, are called the side lengths of the
box. 
The set of boxes with all side lengths between $r$ and $2 r$ is
denoted by $\CalB(r)$.  The set of all boxes in $\RR^d$ will be
denoted by $\CalB$. Lebesgue measure is denoted by $|\cdot|$.

Then Corollary 4.3 of Damanik/Lenz \cite{DL} can be phrased as follows
(see \cite{Len} for related results as well).

\begin{lemma} 
\label{set} Let $\varLambda$ be linearly repetitive. Let $ F : \CalB
  \longrightarrow \RR$ satisfy the following:
\begin{itemize}
\item[(P0)] There exists a $C>0$ such that $|F(B)| \leq C |B|$ for all
boxes with minimal side length at least $1$.
\item[(P1)] There exists a function $b : \CalB\longrightarrow [0,\infty)$ with
  $\lim_{n\to \infty} \frac{b(Q_n)}{|Q_n|}=0$ for any sequence of boxes
  $(Q_n)$ with minimal side length going to infinity such that
$$  F(\cup_{j=1}^n B_j) \leq \sum_{j=1}^n ( F(B_j) + b (B_j)),$$
whenever $\cup_{j=1}^n B_j$ is a box and the $B_j$, $j=1,\ldots,n$, are boxes 
disjoint up to their boundary. 

\item[(P2)] There exists a function $e : [1,\infty)\longrightarrow
[0,\infty)$ with $\lim_{r\to\infty} e(r) =0$ such that $|F(B) - F(x +
B)|\leq e(r) |B|$, whenever $x + B\cap \varLambda = (x+ B) \cap
\varLambda$ and the minimal side length of $B$ is at least $1$.
\end{itemize}

Then, for any sequence of boxes $(Q_n)$ with $Q_n \in \CalB(r_n)$ and
$r_n\to \infty$, the limit $\lim_{n\to \infty} \frac{F(Q_n)}{|Q_n|}$
exists and does not depend on this sequence.
\end{lemma}

\textbf{Remark.} These conditions  have simple
interpretations.  (P1) means that the function $F$ is sub additive up
to a boundary term $b$ and (P2) means that $F$ has an asymptotic
$\varLambda$-invariance property.

\medskip

Let now $\varLambda$ be linearly repetitive. 
As $\varOmega(\varLambda)$ is uniquely ergodic, the autocorrelation
$\gamma =\gamma_\varGamma$ exists for any $\varGamma\in
\varOmega(\varLambda)$. Define the set of local patches
$P(\varGamma)$ of a Delone set $\varGamma$ by
$P(\varGamma):=\{( - x + \varGamma) \cap U_R (0) : R\geq 0, x\in
\varGamma\}.$  Minimality implies 

\begin{equation}\label{liclass}
\varOmega(\varLambda) = \{\Gamma : P(\varGamma)=P(\varLambda)\}.
\end{equation}

In the context of $\RR^d$, we can identify $\xi \in \RR^d$ with the character
$\exp ( i \xi \cdot)$ in $\widehat{\RR^d}$. For $B\subset \RR^d$ relatively
compact with non-empty interior, $\varLambda$ Delone, and $\xi \in \RR^d$, we
define accordingly
$$
C^\xi_{B} (\varLambda):= c^\xi_B (\delta_\varLambda) = \frac{1}{|B|}
\sum_{x\in B\cap \varLambda} \exp (- i \xi x).$$

Now, our result reads as follows. 

\begin{theorem}\label{BTpoint} 
Let $\varLambda$ be linearly repetitive and $\gamma$ the associated
autocorrelation. Then, $$ \gammahat (\{\xi\}) = \lim_{n\to \infty}
|C^\xi_{Q_n} (\varLambda)|^2 $$ for any sequence of boxes $(Q_n)$ with
$Q_n \in \CalB(r_n)$ and $r_n\to \infty$.
\end{theorem}
\begin{proof}Define $F : \CalB\longrightarrow \RR$  by $F(Q) =|\sum_{x\in
  Q\cap \varLambda} \exp(-  i \xi x)|$. Then, $F$ clearly satisfies the
conditions $(P0)$, $(P1)$ and $(P2)$ of Lemma \ref{set}. Therefore, by
Lemma \ref{set}, the limit $\lim_{n\to \infty} \frac{F(Q_n)}{|Q_n|}$
exists for any sequence of cubes $(Q_n)$ with $Q_n \in \CalB(r_n)$ and
$r_n\to \infty$ and the limit does not depend on this sequence.  By
\eqref{liclass}, this means that the limit
$$ a(\xi) :=\lim_{n\to\infty} |C^\xi_{B} (\varGamma)|$$
exists uniformly in $\varGamma \in \varOmega(\varLambda)$ and does not depend
on $\varGamma$. Now, Theorem \ref{abstract}  (a) gives the desired result. 
\end{proof}

\bigskip

We next come to a generalization for arbitrary eigenfunctions.

\begin{prop} \label{ai}
Let $\varLambda$ be a Delone set and $f : \varOmega
(\varLambda)\longrightarrow \CC$ continuous. Then, there exists a
function $ e : [1,\infty)\longrightarrow [0,\infty)$ with $\lim_{r\to
\infty} e(r) =0$ and
$$ \int_B |f(\alpha_{-s} \varGamma) - f(\alpha_{-s} \varGamma')| ds \leq
e(r) |B|$$ whenever $B$ is a box with minimal side length at least $r$
and $\varGamma,\varGamma'\in \varOmega(\varLambda)$ with
$\varGamma\cap B = \varGamma'\cap B$.
\end{prop}
\begin{proof} Define
$e(r)$ to be the supremum of the set of terms
$$\frac{1}{|B|} \int_B |f(\alpha_{-s} \varGamma) - f \alpha_{-s}
\varGamma')| ds,$$ 
where $B$ runs over all boxes  with minimal
side length at least $r$ and $\varGamma,\varGamma'$ belong to
$\varOmega(\varLambda)$ and satisfy $\varGamma\cap B = \varGamma'\cap
B$.

Choose $\varepsilon >0$ arbitrary. As $f$ is continuous, there
exists $R>0$ such that

\begin{equation}\label{kreuzeins}
|f(\varGamma) - f (\varGamma')|\leq \varepsilon
\end{equation}
whenever $\varGamma \cap C_R = \varGamma'\cap C_R$, where $C_R$
denotes the cube centered at the origin with side length $2R$. For a
box $B =[a_1,b_1]\times \ldots [a_d,b_d]$ with minimal side length
bigger than $2R$ set $B_R :=[a_1 -R, b_1-R] \times \ldots[a_d -R, b_d
- R]$. Define $B_R$ accordingly if the intervals making up $B$ are not
closed.  Choose $r_0$ such that
\begin{equation}\label{kreuzzwei}
2 \frac{|B\setminus B_R|}{|B|} \|f\|_\infty \leq \varepsilon
\end{equation}
for any box $B$ with minimal side length at least $r_0$. Then, for such
a box $B$ and $\varGamma$, $\varGamma'\in \varOmega(\varLambda)$ with
$\varGamma\cap B = \varGamma' \cap B$ we have
\begin{equation}\label{kreuzdrei}
|f(\alpha_{-s} \varGamma) - f(\alpha_{-s} \varGamma')| \leq \varepsilon
\end{equation}
for all $s\in B_R$ by \eqref{kreuzeins}. By \eqref{kreuzzwei}, we then
easily infer $e(r) \leq 2 \varepsilon$ whenever $r\geq r_0$. As
$\varepsilon >0$ is arbitrary, this proves the proposition.
\end{proof}

\begin{theorem} Let $\varLambda$ be linearly repetitive. Let $f : \varOmega
(\varLambda)\longrightarrow \CC$ be continuous. Then, for any sequence
of boxes $(Q_n)$ with $Q_n \in \CalB(r_n)$ and $r_n\to \infty$, the sequence
$|\axiq (f)|$ converges uniformly to $ \sqrt{\langle E_T (\{\xi\}) f,
E_T (\{\xi\}) f\rangle} $.
\end{theorem}
\begin{proof} Choose $\Gamma\in \varOmega(\varLambda)$ arbitrary. 
  Define $F : \CalB\longrightarrow \RR$ by $F(Q) =|\int_Q \exp (- i \xi s)
  f(\alpha_{-s} (\varGamma) ds|$.  Clearly, $F$ satisfies $(P0)$ and $(P1)$ of
  Lemma \ref{set}. Moreover, by the previous proposition it also satisfies $
  (P2)$. Thus, the limit $\lim_{n\to \infty} |Q_n|^{-1} F(Q_n)$ exists.
  Another application of the previous proposition and \eqref{liclass} shows
  that the limit does not depend on the choice of $\varGamma$ and is uniform
  in $\varGamma$. Now, the claim follows from Lemma \ref{vNET}.
\end{proof}

We finish this section with a short discussion of linearly repetitive
subshifts. 

Let $\calA$ be a finite set called the alphabet and equipped with the discrete
topology. Let $\varOmega$ be a subshift over $\calA$. Thus, $\varOmega$ is a
closed subset of $\calA^{\ZZ}$, where $\calA^{\ZZ}$ is given the product
topology and $\varOmega$ is invariant under the shift operator
$\alpha:\calA^{\ZZ}\longrightarrow \calA^{\ZZ}$, $(\alpha a) (n)\equiv
a(n+1)$.

We consider sequences over $\calA$ as words and use standard concepts
from the theory of words (\cite{Du,Lot1}).  In particular, $\Sub(w)$
denotes the set of subwords of $w$ and the length $|w|$ of the word
$w=w(1)\ldots w(n)$ is given by $n$.  To $\Omega$ we associate the set
$\Calw=\Calw(\Omega)$ of finite words associated to $\Omega$ given by
$\Calw\equiv \cup_{\omega\in \Omega} \Sub(\omega)$.  A subshift is
called linearly repetitive if there exists a $D>0$ s.t. every $v\in
\Calw$ is a factor of every $w\in\Calw$ with $|w|\geq D |v|$.

A function $F:
\Calw\longrightarrow \RR$ is called subadditive if it satisfies
$F(ab)\leq F(a)+ F(b)$.  If a subshift is linearly repetitive, the
limit $\lim_{|x|\to \infty} \frac{F(x)}{|x|}$ exists for every
subadditive $F$ (see \cite{DL,Len}). This can be used to obtain the
following analogue of the previous theorem.

\begin{theorem} 
Let $\Oomega$ be a linearly repetitive subshift. Let $f$ be a
continuous function of $\varOmega$ and $z \in \CC$ with $ |z|=1$ be
arbitrary. For $n\in \NN$ define the function $A_n (f)$ by $A_n (f)
(\omega) :=\sum_{k=0}^{n-1} z^{-k} f(\alpha_{-k} \omega)$. Then, $|A_n
(f)|$ converge uniformly to the constant function $\sqrt{ \langle E_T
(\{z\})f, E_T (\{z\})f \rangle}$.
\end{theorem}
\begin{proof}  
  Recall that a function $f$ on $\varOmega$ is called locally constant (with
  constant $L\in \NN$) if $f(\omega) = f(\rho)$ whenever $\omega(-L)\ldots
  \omega (L) = \rho(-L)\ldots \rho (L)$. It suffices to show the theorem for
  locally constant functions, as they are dense in the continuous functions.
  To a locally constant function $f$ with constant $L$ we associate the
  function $F :\Calw\longrightarrow \RR$ defined by
\begin{equation*}
F(w) \equiv \left\{\begin{array}{r@{\quad:\quad}l} |w|\|f\|_\infty &
|w|\leq 2 L \\  2L \|f\|_\infty + |\sum_{k=L}^{|w| - L} z^{-k} f( \alpha_k \omega)| &
\mbox{for $\omega\in \varOmega$ with $\omega(1)\ldots \omega(|w|) = w$   if $|w|>2 L$}.
\end{array}\right.
\end{equation*}
As $f$ is locally constant this is well defined. It is not hard to see that
$F$ is subadditive. As discussed above, then the limit $\lim_{|x|\to \infty}
\frac{F(x)}{|x|}$ exists. This easily yields the statements.
\end{proof}

Let us finish this section by emphasizing the following subtle point: As
continuity of eigenfunctions fails for general LR-systems, we can not appeal
to the results of the previous section to obtain validity of $(**)$ for
LR-systems. This does not exclude, however, the possibility that all
eigenfunctions relevant to Bragg peaks are continuous.
We consider this an interesting question.

\subsection*{Acknowledgments}
This work was initiated by discussions with Robert V. Moody at the ``MASCOS
Workshop on Algebraic Dynamics'' in Sydney in 2005. I would like to thank Bob
for generously sharing his insights. I would also like to take the opportunity
to thank the organizers of this workshop for the stimulating atmosphere.  This
work was partially supported by the German Research Council (DFG).

\end{document}